\documentclass{ws-ijmpd}

\newcommand{\beq}{\begin{equation}}
\newcommand{\eeq}{\end{equation}}  
\newcommand{\beqn}{\begin{eqnarray}}
\newcommand{\eeqn}{\end{eqnarray}} 
\newcommand{\gppr}{\stackrel{>}{\scriptstyle \sim}}
\newcommand{\lppr}{\stackrel{<}{\scriptstyle \sim}}

\begin{document}

\markboth{F.M. Rieger}
{Nonthermal Processes in Black-Hole-Jet Magnetospheres}

%
\catchline{}{}{}{}{}
%

\title{Non-thermal Processes in Black-Hole-Jet Magnetospheres}

\author{Frank M. Rieger}

\address{Max-Planck-Institut f\"ur Kernphysik, 69117 Heidelberg, Germany,\\
and European Associated Laboratory for Gamma-Ray Astronomy, jointly supported by CNRS and MPG\\
frank.rieger@mpi-hd.mpg.de}

\maketitle

\begin{history}
\received{11 November 2010}
\revised{18 May 2011}
\end{history}

\begin{abstract} 
The environs of supermassive black holes are among the universe's most extreme phenomena. 
Understanding the physical processes occurring in the vicinity of black holes may provide the 
key to answer a number of fundamental astrophysical questions including the detectability of 
strong gravity effects, the formation and propagation of relativistic jets, the origin of the highest
energy gamma-rays and cosmic-rays, and the nature and evolution of the central engine in Active 
Galactic Nuclei (AGN). As a step towards this direction, this paper reviews some of the progress 
achieved in the field based on observations in the very high energy domain. It particularly focuses 
on non-thermal particle acceleration and emission processes that may occur in the rotating 
magnetospheres originating from accreting, supermassive black hole systems. Topics covered 
include direct electric field acceleration in the black hole's magnetosphere, ultra-high energy 
cosmic ray production, Blandford-Znajek mechanism, centrifugal acceleration and magnetic  
reconnection, along with the relevant efficiency constraints imposed by interactions with matter, 
radiation and fields. By way of application, a detailed discussion of well-known sources 
(Sgr~A*; Cen~A; M87; NGC1399) is presented.
\end{abstract}


\section{INTRODUCTION}
\begin{figure}[t]
\center
\psfig{figure=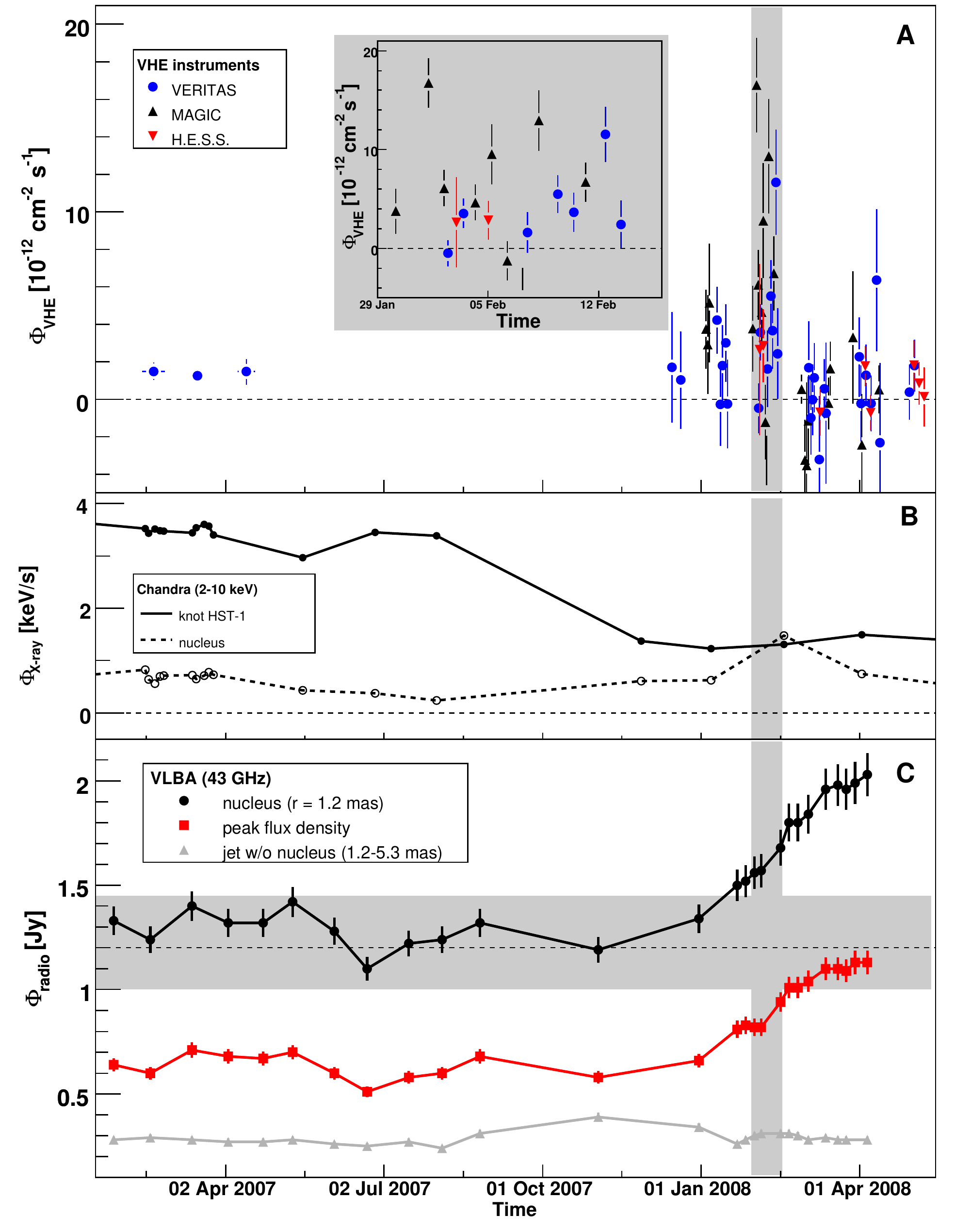, width=10truecm}
\caption{Combined VHE gamma-ray, X-ray and radio light curves for the radio galaxy M87 in 2007--2008. 
A VHE gamma-ray flare (Fig. A) is followed by an increase of the radio flux (Fig. C) close to the black hole. 
Knot HST-1, located in the jet $\sim100$ pc away, is in a low X-ray state (Fig. B) and therefore unlikely to be 
the source of the observed VHE gamma-rays. Details are as follows: (A) VHE -ray flux data ($E > 0.35$ TeV, 
nightly average) including H.E.S.S., MAGIC, and VERITAS. The inlay shows a zoomed version of the rapid 
flaring activity (with timescales as low as $\sim [1-2]$ days) in February 2008. (B) X-ray (Chandra) flux for 
the nucleus and the knot HST-1. (C) Radio (43 GHz VLBA) flux data. The shaded horizontal area indicates 
the range of radio fluxes from the nucleus before the 2008 flare. The radio flux of the outer jet regions does 
not change substantially; most of the observed increase results from the region around the nucleus. Figure 
adapted from Acciari et al.\protect\cite{acciari09}}
\label{m87_lc}
\end{figure}

This review focuses on non-thermal particle acceleration and very high energy (VHE) emission processes 
that may occur in the vicinity of magnetized supermassive black holes. As such, active galaxies and, to 
some extent, extinct or dormant quasars are placed in the center of its attention. Although rotating, 
supermassive black holes have often been considered as putative sources for the energization of ultra-high 
energy (UHE) cosmic rays, non-thermal magnetospheric models have recently gained an additional impetus 
with the detection of very high energy gamma-rays from the (non-blazar) radio galaxy M87, located $\sim 
16$ Mpc away in the Virgo cluster of galaxies.\cite{aharonian06,acciari08,albert08}
While former observations of rapid VHE flux variations (on timescales of $\sim [1-2]$ days) already indicated 
a small size of the $\gamma$-ray emitting region, the location of this region remained ambiguous based on 
the VHE results alone. Additional high-resolution ($\sim 50$ Schwarzschild radii) radio VLBA imaging has 
now shown that the gamma-ray flaring activity is accompanied by an increase in the radio flux close to core 
(see Fig.~\ref{m87_lc}), indicating that the required energetic charged particles may in fact be accelerated 
in the very vicinity of the central black hole.\cite{acciari09} At present, further evidence is needed to strengthen 
this inference. Nevertheless, magnetospheric models where the relevant non-thermal processes are considered 
to occur at the base of a rotating black-hole-jet magnetosphere (e.g., refs.\cite{neronov07,rieger08a,lev11}), 
have emerged as attractive candidates. If verified by further observations, VHE gamma-ray observations could 
provide a fundamental diagnostic of the most violent region in Active Galactic Nuclei (AGN).\\
This review is structured as follows: Sect.~2 gives a phenomenologically-orientated introduction into the 
close environment of supermassive black holes. In Sect.~3, particle acceleration scenarios that draw on 
rotating, large-scale poloidal magnetic fields (i.e., the magnetosphere) are presented. Sect.~4 discusses 
possible interactions of accelerated particles with ambient matter, radiation and fields, that are expected 
to limit achievable particle energies. Applications to concrete sources as potential VHE gamma-ray and
UHE cosmic ray emitters are discussed in Sect.~5.

\section{SUPERMASSIVE BLACK HOLES AND MAGNETIC FIELDS}
\subsection{Evidence and Mass Range}\label{BH_mass}
Most, if not all galaxies, are nowadays believed to harbor a supermassive black hole (BH) with $10^6$ 
to $10^{9.5} M_{\odot}$ at their center\cite{ric98,vika09} (for reviews see also refs.\cite{ferra05,shankar09}). 
Initially hypothesized in order to account for the huge power output seen from extragalactic radio sources 
such as quasars,\cite{zel64} the presence of supermassive black holes has now directly been probed by 
a variety of means, e.g., through the velocity dispersion of stars, water maser line emission or the kinematics 
of nearby ionized gas. The most prominent example is probably the center of our own galaxy where 
near-infrared imaging revealed proper motion of stars which increases with a Kepler law down to separations 
of less than five light days from the compact radio source Sgr~A*, thus providing strong evidence for the 
presence of a black hole of mass $\simeq 3 \times 10^6 M_{\odot}$.\cite{schoedel02,reid09} \\
Several important correlations for the black hole masses in galaxies have been established within recent 
years, including (i) a relation between the black hole mass and the luminosity/mass of its host galaxy bulge, 
e.g., $M_{\rm BH} \simeq 2 \times 10^{-3} M_{\rm bulge}$, and (ii) a relation between the black hole mass 
and the host galaxy stellar velocity dispersion $\sigma$, $\log(M_{\rm BH}/M_{\odot}) = (8.21\pm0.06) +
(3.83\pm0.21) \log(\sigma/200~\mathrm{km/s})$ (with intrinsic scatter of $0.22\pm0.06$ dex).\cite{tundo07} 
For the nearby (z=0.034) gamma-ray blazar Mkn 501 (HBL source), for example, $\sigma=291\pm13$ km/s 
has been found,\cite{falomo02} suggesting a central black hole mass of $\gppr 3\times 10^8 M_{\odot}$. 
Similarly, observations of the giant elliptical radio galaxy NGC 1399 in the centre of the Fornax cluster 
($d \sim 20$ Mpc) suggest $\sigma \simeq 320$ km/s,\cite{houghton06} indicating the presence of a black 
hole with mass exceeding $\sim 5 \times 10^8 M_{\odot}$ (see also ref.\cite{gueltekin09}).

\subsection{Rotating Black Holes}
Supermassive black holes residing in the centers of galaxies may be driven into rotation by prolonged
accretion of angular momentum or as a result of merger events.
\subsubsection{Characteristic properties}
The angular momentum $J$ for a (rotating) Kerr black hole of mass $M$ is usually expressed in terms 
of the dimensionless spin parameter $a\leq 1$, $J=a~J_{\rm max}$, where the maximum value 
$J_{\rm max}$ is given by
\beq
  J_{\rm max} = \frac{GM^2}{c}\,.
\eeq so that an extreme (maximally spinning) Kerr black hole is characterized by a spin parameter 
\beq
   a=\frac{J}{GM^2/c}=\frac{J}{J_{\rm max}} =1\,.
\eeq  For a non-rotating (Schwarzschild) black hole we have $J=a=0$, and the event horizon scale 
$r_H$ coincides with the Schwarzschild radius
\beq
 r_{\rm s} =\frac{2GM}{c^2} \simeq 3 \times 10^{13} \left(\frac{M}{10^8 M_{\odot}}\right)~\mathrm{cm}\,.
\eeq 
\begin{figure}[t]
\centering
\psfig{figure=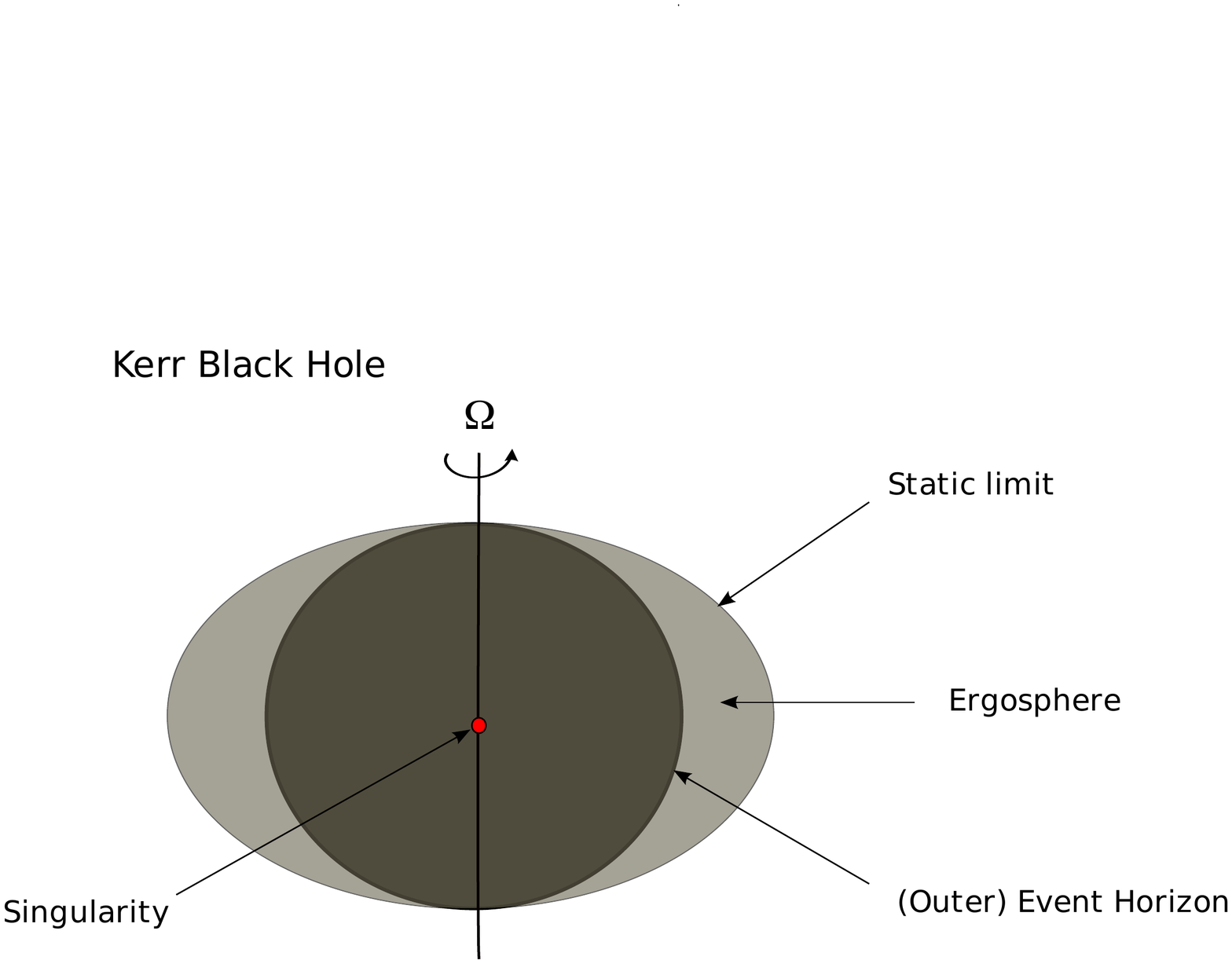, width=10truecm}
\caption{Schematic drawing of the structure of a rotating Kerr black hole.}
\end{figure}
For a rotating Kerr black hole the (spherical) event horizon surface is located at
\beq
r_H = \frac{r_s}{2} \left(1 + \sqrt{1-a^2}\right)\,, 
\eeq with $0\leq a \leq 1$, and where $r_g = r_{\rm s}/2$ is often called the gravitational radius. Hence, 
for a Schwarzschild black hole $r_H = r_{\rm s}$ and for an extreme Kerr black hole $r_H =r_g$. The 
static limit $r_E$ is situated at
\beq
  r_E = \frac{r_{\rm s}}{2}\left(1+\sqrt{1-a^2 \cos^2 \theta}\right)\,
\eeq where $\theta$ is the angle to the polar axis. The region between $r_H < r < r_E$ is called the
ergosphere. Within the ergosphere (which is ellipsoidal in shape), space-time is dragged along in the
direction of the hole's rotation (frame dragging), so that no static observer can exist and a particle has 
to co-rotate with the hole. The angular velocity $\Omega_H$ of a black hole is defined as the angular 
velocity of the dragging of inertial frames at the horizon and given as
\beq\label{Omega_H}
  \Omega_H = a \left(\frac{c}{2r_H}\right)\,,
\eeq so that for an extreme Kerr black hole $\Omega_H = c/2 r_g$.\\ 
A rotating black hole possesses an additional source of energy, which for low spin parameters $a$ is 
given by $E_{\rm rot} =(1/2) I_H \Omega_H^2 = (1/8) a^2 M c^2$ where the black hole's moment of 
inertia $I_H = M r_H^2$ has been employed. For general spins, the maximum rotational energy 
$E_{\rm rot}$ of a rotating black hole can be expressed as the difference between the total mass-energy 
$M c^2$ and its irreducible mass $M_{\rm irr} c^2$, see refs.\cite{christ70,thorne86}, 
\beq\label{maxErot}
  E_{\rm rot} = (M-M_{\rm irr}) c^2 = M c^2 \left(1-\sqrt{0.5~(1+\sqrt{1-a^2})}\right) \leq 0.29 M c^2\,.
\eeq Thus, if supermassive black holes would indeed be rapidly rotating and if they would be able
to dissipate all of their rotational energy over a characteristic life time $t_{\rm Edd} = 4.5 \times 10^7$ 
yr (see Eq.~[\ref{edd_time}]), this could yield a considerable luminosity output 
\beq\label{maxLrot}
 L_{\rm rot} = \frac{E_{\rm rot}}{t_{\rm Edd}} \leq 3.6 \times 10^{46} 
                        \left(\frac{M}{10^8 M_{\odot}}\right)~\mathrm{erg/s}\,,
\eeq comparable in magnitude to the Eddington luminosity (see Eq.~[\ref{ledd}]). The maximum output 
is, however, very sensitive to the spin parameter $a$. For moderate spins, e.g., $a = 0.5$, it is already 
about an order of magnitude smaller ($E_{\rm rot} =0.034 M c^2$), see Fig.~(\ref{maxErot_fig}) for 
illustration.
\begin{figure}[t]
\center
\psfig{figure=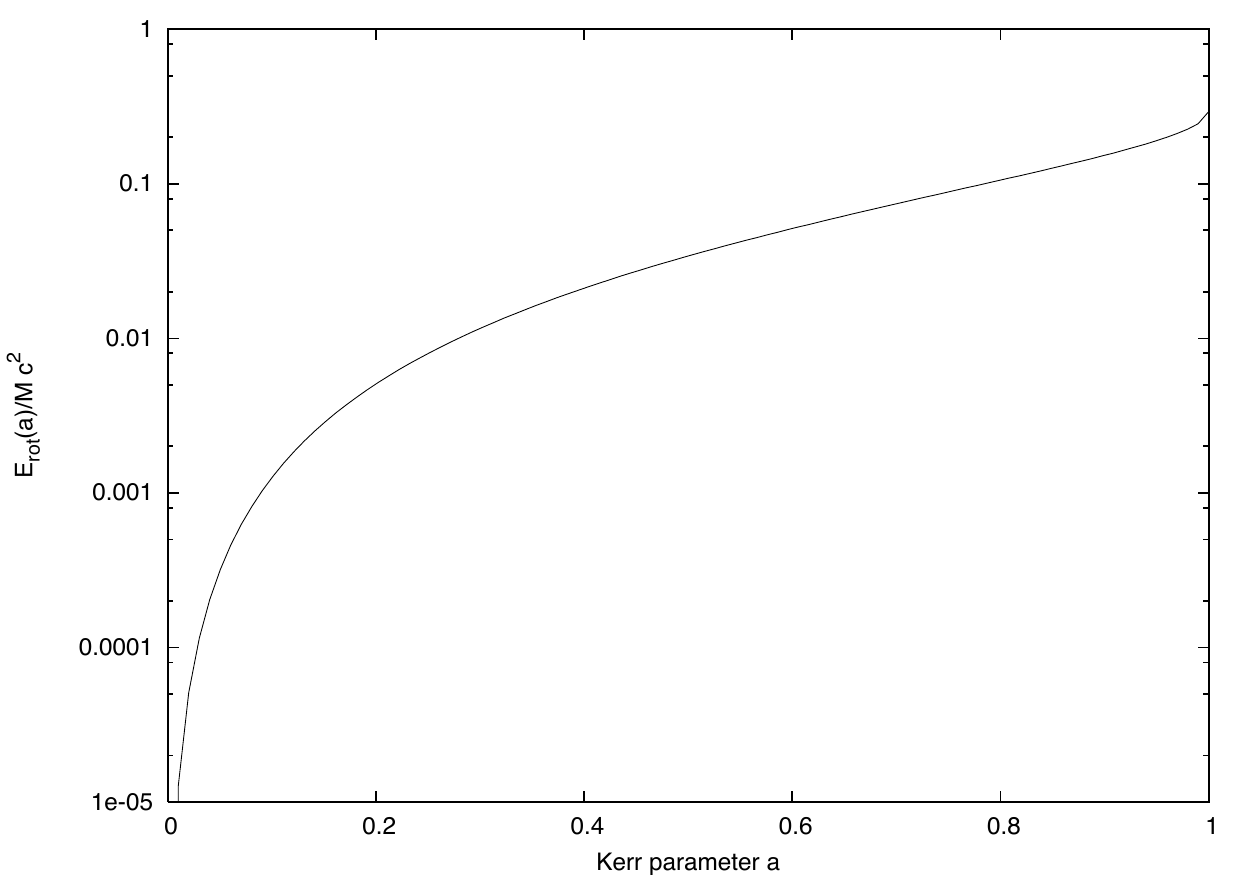, width=10truecm}
\caption{Maximum extractable energy of rotating black hole as a function of the dimensionless spin 
parameter $a$.} 
\label{maxErot_fig}
\end{figure}

\noindent
Note that the most general, stationary black hole metric is the Kerr-Newman metric, where the black 
hole is described by its mass $M$, electric charge $Q$ and spin angular momentum $J$. Special cases 
are the Kerr ($Q=0, J \neq 0$), the Reissner-Nordstr\"om ($J=0, Q\neq 0$) and the Schwarzschild ($J=
Q=0$) black hole. Charged black holes are commonly thought not to be of astrophysical importance 
(but see also, e.g., refs.\cite{punsly98,ruffini10}), as the surrounding plasma is usually expected 
to rapidly neutralize any charge imbalance (e.g., refs.\cite{eardly75,novikov89}).\footnote{The metric 
function implies the constraint $(J/M)^2+ (GQ^2/c^2) \leq (GM/c)^2$, so that the maximum, theoretically 
possible charge would be $Q_m=G^{1/2} M$. However, real astrophysical black holes are expected to 
possess only negligible charge (i.e., $Q \lppr G M m_p/e \sim 10^{-18} Q_m$) as otherwise electric forces 
would dominate over gravitational ones, enabling charge separation and attracting opposite charge to 
neutralize the imbalance. This implies that the magnetic field produced by a charged black hole is limited 
to $B\lppr m_p c^2/(e r_g) \sim 2 \times 10^{-7} (10^8 M_{\odot}/M)$ G.}

\subsubsection{On the Distribution of Black Hole Spins}
Despite their obvious significance, the spins of astrophysical (supermassive) black holes have been 
proven difficult to pin down unambiguously. This may appear somewhat unfortunate, as it e.g. directly 
affects inferences about the maximum rotational energy that could possibly be extracted by some 
electrodynamic processes (cf. Eq.~[\ref{maxErot}]). Naively, one expects that supermassive black holes 
can be spun-up by prolonged accretion of angular momentum or by the merger of black holes with similar 
masses (having high orbital angular momentum). On the other hand, if a rotating black hole would be 
able to emit a strong (persistent) spin-powered jet of, e.g., $L_j \sim 10^{46} (M/10^8M_{\odot})$ erg/s, 
the hole's rotation could spin down on an e-folding timescale $E_{\rm rot}/L_j \sim 10^8$ yr comparable 
to the typical AGN lifetime. Then again, prolonged accretion or further merger events could help to sustain 
high black hole spins. Prolonged accretion is expected to lead to high spins ($a\sim 1$),\cite{volonteri05} 
whereas major mergers tend to produce average spins of $\langle a \rangle \sim 0.7$.\cite{berti08} 
Provided accretion is not inherently random and episodic, cosmological N-body simulations suggest that 
the final black hole spin depends almost exclusively on the accretion history and very little on the merger 
history.\cite{lagos09} Accordingly, more massive black holes (i.e., those hosted by elliptical galaxies) 
should be characterized by higher spin parameters (average $\langle a \rangle \geq 0.8$ at $z=0$) 
compared to those residing in, e.g. spirals. However, the underlying assumption of a spin evolution via 
prolonged accretion has been questioned recently.\cite{king08} In the standard picture, Lense-Thirring 
precession is assumed to always ensure co-alignment of the disk with the black hole spin, and thus to 
lead to spin-up via prograde accretion.\cite{bardeen75} However, this needs not necessarily to be the 
case if the disk is sufficiently small. Instead, accretion may then be random and episodic in character, 
so that the spins are expected to readily adjust to average values $\langle a \rangle \sim 0.1-0.3$. Such 
a spin-down scenario seems in fact supported by recent studies of the radiative efficiency from accretion 
onto supermassive black holes:\footnote{The radiative efficiency $\eta$ is usually defined as $\eta=
1-e_{\rm ms}/c^2$, where $e_{\rm ms}=c^2(1-r_s/3r_{\rm ms})^{1/2}$ is the binding or specific energy of 
a test particle rotating in the innermost stable circular orbit at $r_{\rm ms}$, and becomes $\eta = 0.057$ 
for a Schwarzschild and $\eta=0.42$ for an extreme Kerr black hole ($a=1$). Note that while accretion 
by itself would spin up the black hole rather quickly to its maximum value $a=1$, photon capture will 
prevent complete spin-up and limit the maximum spin to $a=0.998$. While this is still very close to $a=1$, 
it has a significant effect on the maximum possible radiative efficiency, reducing it to $\eta=0.3$.\cite{thorne74}} 

\noindent (i) Existing quasar survey data suggest a significant decrease of the radiative efficiency from high 
($\eta \sim 0.2-0.3$ at $z\simeq 2$) to low redshift ($\eta \sim 0.03$ at $z=0$). While the spin evolution at 
$z \sim 2$ may be driven by major merger events, the decrease of the radiative efficiency towards low $z$ 
seems to support an interpretation according to which black holes are fed by accretion events that tend to 
spin down the black hole over time ($a\sim 0$ at $z=0$).\cite{wang09} 
(ii) Similarly, a rather low mean efficiency $\langle \eta \rangle \sim 0.07$ has been inferred based on the 
cumulative energy density emitted by AGNs over the age of the Universe. The results indicate $\langle a 
\rangle \simeq (0.2-0.6)$ and suggest that supermassive black holes are on average not rapidly spinning 
during accretion.\cite{martinez09} Note that if Maxwell stresses do (contrary to what is usually assumed) 
not vanish at $r_{\rm ms}$, as in fact seen in recent MHD disk simulations,\cite{hawley06} this could lead 
to an excess dissipation, lowering the above inferred spins even more (possible differences seem to be 
small, though, at least in the case of a thin disk\cite{noble09}).

\noindent Yet, while on average rather moderate values of $a$ seem to be preferred, a sub-population 
of AGNs with high spins may well exist. In fact,  X-ray observations of relativistically broad iron K$\alpha$ 
lines from the black hole's vicinity, e.g., in the prominent Seyfert 1 galaxy MCG-6-30-15, for which $a>0.9$ 
has been inferred, indicate that at least some radio-quiet (!) AGN may harbor rapidly spinning black holes 
(for review, see refs.\cite{reynolds03,brenneman09}).\footnote{Note however, that recent modelling of the broad 
iron lines in the Seyfert-1 AGNs Fairall 9 and Swift J2127.4+5654 seems to favor more moderate spin values 
$a \sim 0.6$, making MCG-6-30-15 appear somewhat exceptional.\cite{schmoll09,miniutti09}} If indeed 
the case, then this implies that a high black hole spin is not a sufficient condition for radio-loudness as initially 
suggested by Wilson \& Colbert\cite{wilson95}. On the other hand, evidence for rather low black hole spins 
in the galactic microquasar sources XTE J1550-564 ($a\simeq 0.1$) and A0620-00 ($a\simeq 0.3$) also 
challenge the common perception that black hole spins and relativistic jets are necessarily 
connected.\cite{mcclintock11} Fender et al.\cite{fender10} have recently concluded that at least for galactic 
black hole X-ray binaries (which are often believed to be scaled-down versions of AGNs), there is no 
evidence for either the jet power or the jet speed being related to the spin of the black hole. If this is indeed 
the case, it may point to the fact that the jets rather emerge as centrifugally-driven (Blandford-Payne) 
outflows from the accretion disk. 

\noindent With this caveat in mind, estimates for the jet kinetic power (e.g., from intracluster X-ray cavities 
inflated by the radio jets) and for the central black hole mass have been employed to constrain the spin
parameters for selected samples of radio galaxies. Some of these studies conclude that the radio-loud 
dichotomy of FR~I and FR~II radio galaxies can be reproduced by a transition in accretion mode from a 
standard disk to an ADAF, with the black holes for both classes rapidly rotating ($a \gppr 
0.9$).\cite{wu08,wu11}
 Others instead find that the FR~I type and FR~II sources they studied are
consistent with spin values of $a \simeq (0.01-0.4)$ and $a \simeq (0.2-1)$, respectively, and report 
evidence for a decrease of the black hole spin in FR~II with decreasing redshift.\cite{daly11}  
At present, further studies are certainly needed to clarify this issue.

\subsection{Accretion onto Black Holes}
\subsubsection{Spherical  accretion and Eddington constraints}
The luminosity of an accretion-powered object of mass $M$ cannot grow indefinitely. In the case of 
steady spherical accretion, the maximum luminosity is limited to the Eddington value $L_{\rm Edd}$ 
given by the balance between (outward-directed) radiation-pressure and (inward-directed) 
gravitational force,
\beq\label{ledd}
L_{\rm Edd}=\frac{4\pi c G M m_p}{\sigma_T}=1.25 \times 10^{46} \left(\frac{M}{10^8M_{\odot}}\right) 
                        \mathrm{erg/s}\,.
\eeq The critical accretion rate required to sustain the Eddington luminosity, assuming a typical 
$\eta=0.1$ efficiency of conversion of mass to radiant energy, is then defined as
\beq
  \dot{M}_{\rm Edd}=\frac{L_{\rm Edd}}{\eta c^2} 
                                    \simeq 1.4 \times 10^{26} \left(\frac{M}{10^8M_{\odot}}\right)~\mathrm{g/s} 
                                    \simeq 2.2 \left(\frac{M}{10^8M_{\odot}}\right)~M_{\odot}~\mathrm{yr}^{-1}\,.
\eeq 
The Eddington timescale (sometimes called the Salpeter timescale) is the e-folding growth 
time for the mass of a black hole accreting at the Eddington rate
\beq\label{edd_time}
 t_{\rm Edd} = \frac{M}{\dot{M}_{\rm Edd}}=4.5 \times 10^7 \mathrm{yr}\,.
\eeq The fiducial Eddington magnetic field upper limit $B_{\rm Edd}$ has an energy density equal 
to the radiative energy density at $r_g$ of a body emitting at $L_{\rm Edd}$, i.e., $B=(2 L_{\rm Edd}/
r_g^2 c)^{1/2}$, 
\beq\label{Bedd}
  B_{\rm Edd} \simeq 6.1 \times 10^4 \left(\frac{10^8 M_{\odot}}{M}\right)^{1/2}~ \mathrm{G}\,. 
\eeq

\subsubsection{Standard, steady state accretion flows}\label{standard_disk}
\begin{figure}[ht]
\centering
\psfig{figure=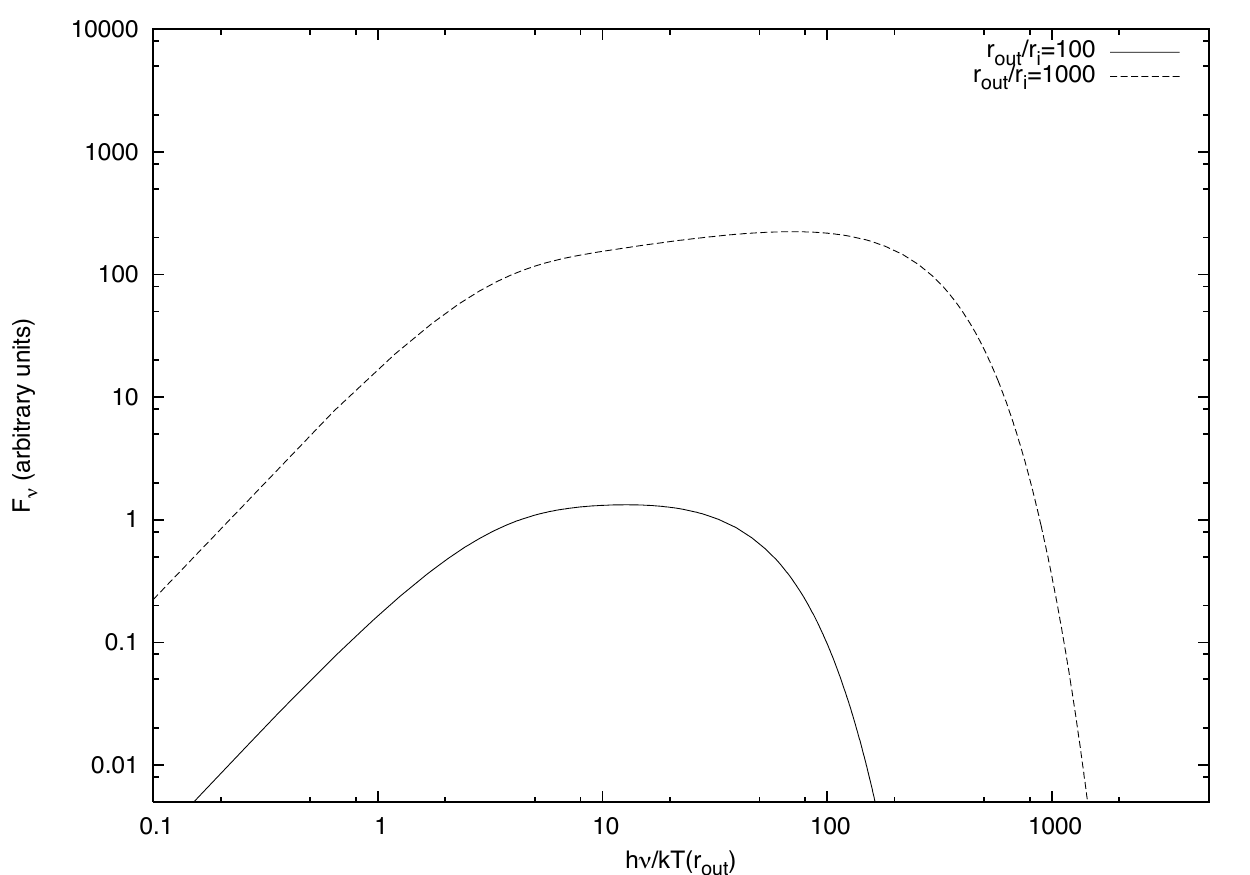, width=10truecm}
\caption{The characteristic continuum spectrum of a standard, geometrically thin, optically thick disk 
for two different outer radii $r_{\rm out}$. The spectrum is normalized to $h\nu/kT(r_{\rm out})$. If the 
disk is not large enough, e.g., for $r_{\rm out}/r_{\rm in} =100$, the $\nu^{1/3}$--power law part may not 
become apparent.} 
\label{disk_sed}
\end{figure}

In general, the structure of an accretion flow is determined by the balance between gravitational 
heating and cooling. Its concrete structure therefore depends on what kind of heating and cooling 
processes are assumed to be dominating. In the classical approach of Shakura \& Sunyaev\cite{shakura73}, 
commonly referred to as the {\it standard accretion disk model}, the disk is flat (geometrically thin: 
$H\ll r$; $H$ the half thickness of the disk, $r$ the radial distance) and opaque (optically thick in 
vertical direction). Viscous (frictional) stresses\footnote{With the r$\phi$-component of the shear 
stress tensor taken to be proportional to the total (gas and radiation) pressure $p$, i.e., $t_{r\phi}=
\alpha p$, $\alpha \leq 1$ ($\alpha$-prescription).} are assumed to convert gravitational potential 
energy into heat that is released locally in the form of a thermal black body spectrum $B_{\rm \nu}
(T_{\rm eff}[r])$ (i.e., possible modifications by advection or a jet are neglected) with energy flux per 
unit surface area of $F(r)=\pi \int B_{\nu}(T_{\rm eff}[r]) d\nu=\sigma T_{\rm eff}(r)^4$ and characteristic 
effective temperature 
\beqn\label{ss_temperature}
T_{\rm eff}(r) &=&\left[ \frac{3GM\dot{M}}{8\pi\sigma r^3}\left(1-\sqrt{\frac{r_{\rm in}}{r}}\right)\right]^{1/4}
                                \nonumber \\
                       &= & 6.3 \times 10^5 \left(\frac{\dot{M}}{\dot{M}_{\rm Edd}}\right)^{1/4}
                                       \left(\frac{10^8 M_{\odot}}{M}\right)^{1/4}
                                       \left(\frac{r_s}{r}\right)^{3/4}\left(1-\sqrt{\frac{r_{\rm in}}{r}}\right)^{1/4}
                                       \mathrm{K}\,.
\eeqn Far from the inner edge, $r \gg r_{\rm in}$, the temperature approximately obeys $T_{\rm eff}(r) 
\propto r^{-3/4}$. As gravity is stronger on smaller scales, the disk surface is hotter in the inner region. 
The radially integrated disk luminosity is 
\beq\label{ss_disk}
L_ d =2~\int_{r_{\rm in}}^{r_{\rm out}} 2\pi r F(r)~dr \simeq \frac{GM\dot{M}}{2r_{\rm in}}
         = \frac{1}{4} \left(\frac{r_s}{r_{\rm in}}\right)\dot{M} c^2\,. 
\eeq Half of the total gravitational energy release $GM\dot{M}/r_{\rm in}$ is thus radiated away, the 
remaining half being accounted for by the kinetic energy of the orbiting material. The emerging disk 
spectrum $F_{\nu} \propto \int B_{\nu}(T_{\rm eff}[r]) r dr$ can be obtained by integrating $B_{\rm \nu}$ 
over the surface of the disk. The resultant spectrum is shown in Fig.~\ref{disk_sed}: It initially rises 
with $\nu^2$, then flattens (provided the disk is large enough) in the intermediate region to $\nu^{1/3}$, 
before it decays exponentially towards the highest frequencies.  
For characteristic AGN parameters, the emission from the inner part is maximized at a frequency 
\beq\label{ss_disk_frequency}
\nu_{\rm max}=\frac{2.82 k T_{\rm eff}}{h}\simeq 5.9 \times 10^{15} 
                            \left(\frac{T_{\rm eff}}{10^5~\mathrm{K}}\right) \mathrm{Hz}\,,
\eeq suggesting that this type of accretion flow could lead to a 'big blue bump' feature, i.e. a significant
optical/UV disk contribution to the observed SED.\\ 
As a consequence of the viscous interactions between adjacent layers in the disk, angular momentum 
is transferred outwards, while mass falls (accretes) inwards.  

\subsubsection{Inner edge of the accretion disk}
In the classical picture, the radius of the inner edge $r_{\rm in}$ of the standard disk extends down to 
the (innermost) marginally stable circular orbit $r_{\rm in} = r_{\rm ms}$. Inside $r_{\rm ms}$ space-time 
gets so strongly curved that stable orbits no longer exist. A particle thus unavoidably falls into the black 
hole. This is why the region inside $r_{\rm ms}$ is often called the "plunging region" as gas can spiral 
into the black hole without further loss of angular momentum. The analytical solution for the innermost 
stable orbit as a function of the dimensionless spin parameter $a$ is given by\cite{bardeen72} 
\beq
  r_{\rm ms} = \frac{r_s}{2} \left[3+Z_2 \mp \sqrt{(3-Z_1)(3+Z_1+2Z_2)}\right]
\eeq where 
\beqn
Z_1 &=& 1+(1-a^2)^{1/3} \left[(1+a)^{1/3}+(1-a)^{1/3}\right]\nonumber\\
Z_2 &=&\sqrt{3a^2+Z_1^2}\,.
\eeqn Hence, for a Schwarzschild black hole ($a=0$), $r_{\rm ms} = 3 r_s = 6 r_g$, while for the extreme
Kerr case ($a=1$), $r_{\rm ms} = r_s/2 = r_g$ (prograde orbit) or $r_{\rm ms} = 9 r_s/2 = 9 r_g$(retrograde 
orbit). As mentioned above, the disk's radiative efficiency is strongly dependent on the position of its 
innermost radius.

\subsubsection{Radiatively inefficient accretion flows}\label{riaf}
\begin{figure}[ht]
\vspace*{-0.5cm}
\centering
\psfig{figure=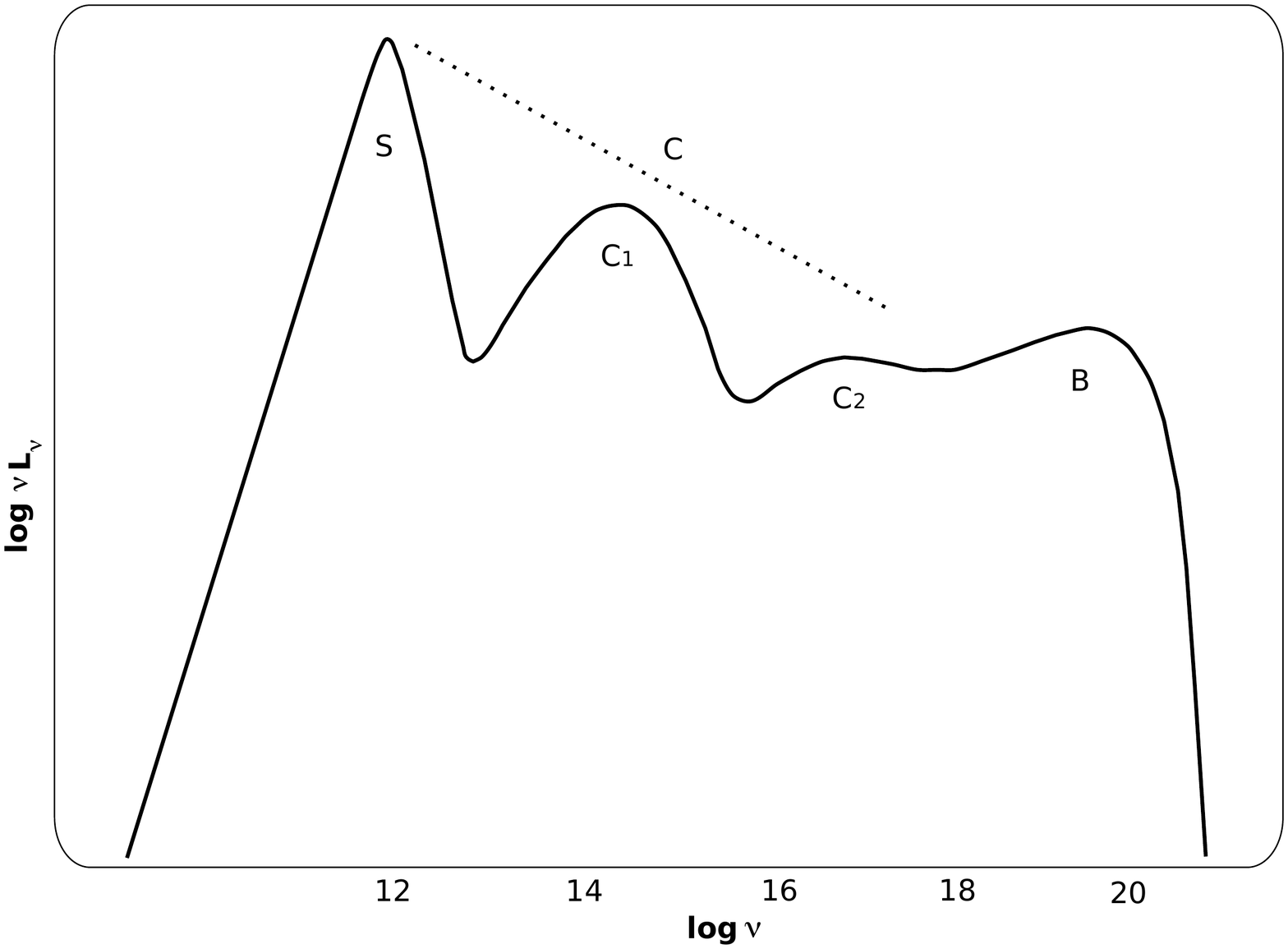, width=12truecm}
\caption{Sketch of a typical ADAF spectrum around a supermassive black hole. Synchrotron 
emission (S) of the relativistic thermal electrons produces a peak at around $10^{12}$ Hz. 
At moderate accretion rates, Comptonization results in a power-law tail (C) extending into 
the X-ray domain. For low accretion rates, distinct Compton (C1=once Compton scattered, 
C2=twice Compton scattered) peaks appear. The X-ray/soft gamma-ray regime is usually 
dominated by bremsstrahlung emission (B), peaking at $h\nu \simeq k T_e$, and with 
luminosity scaling as $L_{\rm br} \propto \dot{m}^2$.}
\label{adaf_sed}
\end{figure}

In the standard approach, cooling is so efficient that all the energy released through viscosity is radiated
away locally. This may not always be the case and in fact we know steady disk solutions where this 
assumption is violated (for a review, see Kato et al.\cite{kato08}). In the case of a radiatively inefficient 
accretion flow (RIAF), for example, most of the energy released is stored within the flow and transported 
inward with accretion. While the disk thus locally undergoes advective cooling, the flow itself becomes very 
hot. When the density of the accretion flow falls below a critical value, the flow becomes optically thin and 
the efficiency for radiative cooling (e.g., bremsstrahlung, which dominates at non-relativistic temperatures) 
becomes small. The classical RIAF prototype is the two-temperature, optically-thin advection-dominated 
accretion flow (ADAF) model (for review, see refs.\cite{narayan98,yi99}). By assumption, the 
released viscous energy is assumed to go primarily into ion heating, while cooling is mainly by the 
electrons. If the densities are sufficiently low, Coulomb coupling between protons and electron is weak 
and the viscous energy transfer from protons to electrons becomes very small, limiting the amount of 
energy that can be lost by radiation. This introduces a critical accretion rate 
\beq\label{mcrit}
\dot{M} \sim \alpha^2 \dot{M}_{\rm Edd}
\eeq above which an ADAF cannot exist (typically, $\alpha \sim 0.2-0.3$). Because of the very high 
temperature ($T_p \sim GMm_p/3k_B r  \sim 3 \times 10^{12} (r_g/r)$ K for ions, $T_e \sim 5\times 
10^9$ K for electrons) ADAFs are marginally geometrically thick with scale height $H/r  \sim 1$ at every 
radius, independent of the accretion rate. The total radiative luminosity of an ADAF is found to be 
proportional to the square of the mass accretion rate, in contrast to the standard thin disk where $L 
\propto \dot{M}$ (eq.~[\ref{ss_disk}]), i.e. 
\beq
L_{\rm ADAF} \sim \frac{2 \times 10^{-2}}{\alpha^2} \left(\frac{\dot{M}}{\dot{M}_{\rm Edd}}\right)\dot{M} c^2 
                         \propto \dot{M}^2\,
\eeq (Mahadevan 1997), and thus can be much smaller than the luminosity expected from a 
standard disk. The broadband spectrum emerging from an ADAF can cover the energy range from 
radio to hard X-ray frequencies, see Fig.~\ref{adaf_sed}. The radio to sub-mm regime is typically 
produced by synchrotron emission of relativistic thermal electrons and in its optically thin part rises 
with $L_{\nu} \propto \nu^{2/5}$.\cite{mahadevan97} The emission at the highest peak frequency 
$\nu_p$ comes from the innermost part of the disk\cite{mahadevan97,yi99}
\beq\label{ADAF_peak}
 \nu_p(r) \simeq 10^{12}~ \left(\frac{10^8 M_{\odot}}{M}\right)^{1/2}
                \left(\frac{\dot{M}}{0.01 \dot{M}_{\rm Edd}}\right)^{1/2} \left(\frac{T_e}{10^9\mathrm{K}}\right)^2
                \left(\frac{r_s}{r}\right)^{5/4}\,~\mathrm{Hz}\,,
\eeq while the lower frequency synchrotron emission originates from further out. Compton upscattering 
of the low-energy synchrotron photons by the relativistic thermal disk electrons results in a hard power 
law tail extending up to energies of $h \nu \sim k T_e \sim (100-500)$ keV. Compton scattering becomes 
less important with decreasing accretion rates, and the spectrum becomes steeper (softer). For sufficiently 
low $\dot{M}$ distinct Compton peaks can appear. This is related to the increase in electron temperature 
with decreasing accretion rates. For low $\dot{M}$, the X-ray spectrum is dominated by bremsstrahlung 
emission (with a non-negligible contribution from large radii) due to electron-electron and electron-ion 
interactions, again up to a maximum energy of $\sim kT_e$, beyond which the spectrum falls off 
exponentially.\cite{manmoto97}

\noindent A number of additional scenarios have been discussed in the literature that extend and 
complement the classical ADAF picture described above. One important modification concerns, 
e.g., jets and winds. Since for self-similar ADAFs the net energy of the accretion flow can remain 
positive and unbound to the central object (positive Bernoulli number),\cite{narayan94} strong 
outflows may well form and modify the emergent spectra, cf. the ADIOS model by Blandford \& 
Begelman\cite{blandford99}. While the positiveness of the Bernoulli number is a necessary (but 
not sufficient) condition for outflows, 2-D simulations indeed suggest that for large viscosity 
parameters ($\alpha \gppr 0.3$) strong outflows can occur.\cite{igum00,abramow00} If mass 
is supplied by an outer standard disk, ADIOS-like outflows may however only occur for 
transition radii $r_t \gppr 100 r_s$.\cite{turolla00}

\subsection{Magnetic Field Strengths}\label{mfields}
In the standard disk case, we can derive a characteristic (maximum local equipartition) magnetic field 
strength close to $r_g$, whose energy density is comparable to that of the radiation (i.e., $B^2/8\pi =
\sigma T_{\rm eff}^4/c$) which gives (cf. eq.~[\ref{ss_temperature}]) 
\beq\label{mfield_rad}
  B_r \simeq 4.4 \times 10^4  \left(\frac{\dot{M}}{\dot{M}_{\rm Edd}}\right)^{1/2}
                                                    \left(\frac{10^8 M_{\odot}}{M}\right)^{1/2} \mathrm{G}\,.
\eeq For $\dot{M}=\dot{M}_{\rm Edd}$ this is of about the same strength as the Eddington equipartition 
magnetic field $B_{\rm Edd}$, eq.~(\ref{Bedd}), obtained by assuming that all of $L_{\rm Edd}$ emerges 
from $r_g$.  In the case of an advection-dominated accretion flow (ADAF) on the other hand, the 
characteristic equipartition magnetic field is given by $B_a^2/8 \pi \simeq 0.5 \rho c_s^2$. Here $c_s$ 
denotes the isothermal sound speed, $\rho \propto \dot{M}/(r^2v_r)$ is the density of accreted matter, 
$v_r \sim 0.5 \alpha v_f$ is the typical radial infall speed and $v_f =(GM/r)^{1/2}$ the free fall velocity. 
When scaled to the AGN mass range, this becomes\cite{narayan98} 
\beq\label{mfield_adaf}
 B_a\simeq 7.8 \times 10^4 ~\alpha^{-1/2} \left(\frac{\dot{M}}{\dot{M}_{\rm Edd}}\right)^{1/2}
                      \left(\frac{10^8 M_{\odot}}{M}\right)^{1/2} \left(\frac{r}{r_s}\right)^{-5/4}
                      \mathrm{G}\,.
\eeq As an ADAF cannot exist above $\dot{M} \sim \alpha^2 \dot{M}_{\rm Edd}$, cf. eq.~(\ref{mcrit}), the 
maximum $B_a$ is comparable to the maximum $B_r$ in the standard case.\\ 
Note that the poloidal field strength in the magnetosphere may be significantly smaller than these disk 
field values. Taking $B_r$ to be characteristic for the poloidal magnetospheric field component, for 
example, may then well overestimate the possible power of a Blandford-Znajek-type process. It has in 
fact been argued,\cite{livio99}  that the large-scale poloidal field $B_{\rm pd}$ threading the disk should 
be related to the small scale field produced by dynamo processes via $B_{\rm pd} \sim (H/r) 
B_{\rm dynamo}$, where $H/r$ is the typical disk scale height, which for a standard disk is much smaller 
than unity (cf. however also ref.\cite{meier02}). In any case, as the field threading the hole is generated by 
currents in the disk, it seems unlikely that its strength should significantly exceed the field threading the 
inner disk.

\subsection{Black Hole-Jet-Magnetospheres}\label{magnetosphere}
\begin{figure}[t]
{\psfig{figure=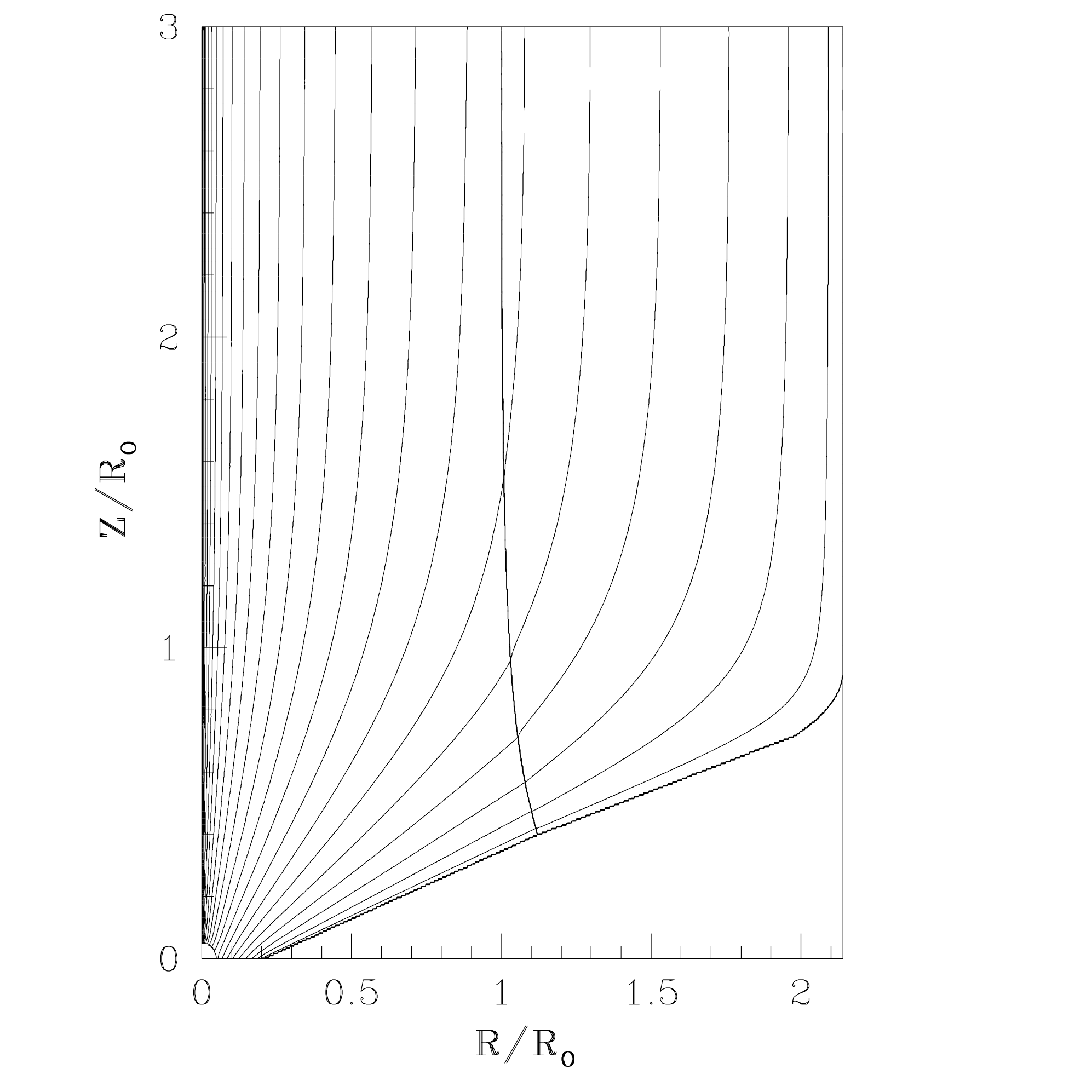, width=7truecm}}{\psfig{figure=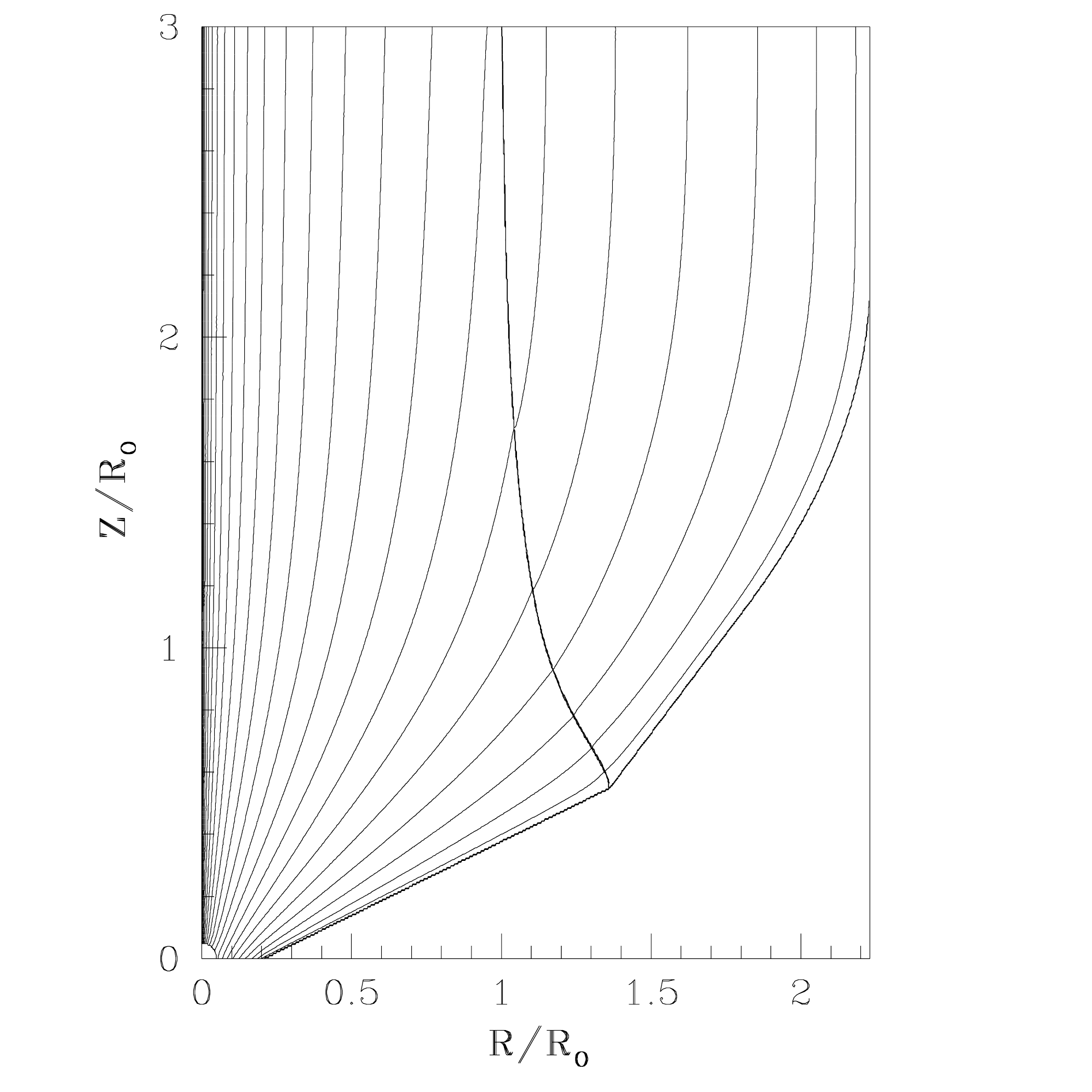, width=7truecm}}
\caption{Poloidal magnetic field structure of a collimating, relativistic jet, obtained from the numerical solution of the 
stationary-state, axisymmetric relativistic MHD equations. The magnetosphere is assumed to be force-free. Dipolar 
magnetic flux carried by the accretion disk is stretched by a disk wind into the upper hemisphere. Left: Assuming 
weak  differential (quasi-rigid) rotation of foot points of the field lines. Right: For strong differential rotation. 
$R_0$ denotes the asymptotic light cylinder radius. From Fendt and Memola\protect\cite{fendt01}.
}\label{fendt}
\end{figure}

Once a spinning black hole or disk becomes threatened by an ordered (large-scale poloidal) magnetic field, it can 
build up a rotating magnetosphere filled up with currents. The required magnetic flux could be advected from the
interstellar magnetic field via the accretion process and/or be produced (amplified) through dynamo actions in the
inner accretion disk.\footnote{An alternative scenario relates to the Poynting-Robertson radiation force acting 
predominantly on the electrons in the accretion disk. As a consequence, electrons lag behind the protons, and
a toroidal electric current develops that can become sufficiently large to produce significant poloidal magnetic 
fields.\cite{contopoulos06}} While this picture may appear qualitatively evident, the detailed electromagnetic 
structure of  the magnetosphere is a highly complex problem. 
Most research has been influenced by the seminal papers of Blandford \& Znajek (BZ)\cite{blandford77} and 
Blandford \& Payne\cite{blandford82}. 
Roughly speaking, the main difference concerns the question whether the source of rotational energy and 
angular momentum lies with the black hole's rotation or with the accretion disk. Methodologically, this often
translates into two different approaches (for review and further references, see e.g. refs.\cite{blandford02,camenzind07}): \\
In the first, the {\it force-free electrodynamic} (FFE) approximation, the magnetospheric structure is assumed to be 
governed by Ampere's and Faraday's law and to be force-free everywhere (i.e., dominated only by magnetic and 
electric stresses). Accordingly, gravitational and inertial forces are neglected (massless limit of MHD) and the 
current density $\vec{j}$ perpendicular to the local magnetic field is determined from the force-free condition 
$\rho \vec{E}_{\perp}+\vec{j}\times \vec{B}/c=0$, where $\rho$ is the charge density.\cite{komissarov02} This 
approximation quite simplifies the standard problem in that it allows to solve for the electromagnetic field structure 
without having to solve for the plasma dynamics. Magnetic field lines may then be thought as (quasi-rigidly) rotating 
with the angular velocity $\Omega$ of the horizon or the inner disk (Ferraro's law). Approaching the light surface 
where $|\vec{\Omega} \times \vec{r}| =c$ the field is swept backwards opposite to the sense of rotation and a toroidal 
twist is introduced. By assumption, plasma in the magnetosphere is streaming outward at almost the speed of light,
its only role being to provide the currents required to sustain the magnetic field.\\
In the second approximation, the {\it relativistic MHD (single fluid) approach}, the force-free condition is modified to 
allow for inertial and pressure terms in the equation of motion. If plasma loading is not negligible, as may be the case 
for field lines connecting to the disk, such a modification appears unavoidable. The system is then closed by Ohm's law 
asserting that the current density $\vec{j}$ is proportional to the electric field $\vec{E}'=\vec{E}+\vec{v}\times \vec{B}/c$ 
seen in the frame where the fluid is at rest. In the high conductivity (ideal MHD) limit, $\sigma_e \rightarrow \infty$, 
Ohm's law implies $\vec{j}/\sigma_e = \vec{E}+\vec{v}\times \vec{B}/c \rightarrow 0$ where $\vec{v}$ is the fluid velocity. 
Within the Alfv\'{e}n surface\footnote{Defined in the non-relativistic case as the surface where the plasma kinetic energy 
density equals the magnetic field energy density. Relativistic flows remain Poynting-dominated at the Alfv\'{e}n surface,
with the Alfv\'{e}n surface only being related to the propagation of Alfv\'{e}n waves, that become static at this location.} 
plasma that becomes attached to the field lines behaves like a bead on a wire and is flung out by the centrifugal force 
given suitable field line orientations (e.g., inclinations wrt to disk launching of less than $60^{\circ}$ in Newtonian 
approximation, and nearly $90^{\circ}$ in the extreme Kerr limit.\cite{cao97,lyutikov09,sadowski10}).
For stationary and axisymmetric configurations, the equation of motion (which in the low inertia limit just reduces 
to the force-free condition) projected perpendicular to the field lines gives the famous Grad-Shafranov equilibrium 
(ideal MHD) equation. The so-derived Grad-Shafranov equation (GSE) is a highly non-linear partial differential 
equation and (while simply re-expressing the cross-field force balance) highlights that the structure of the 
magnetosphere essentially depends on the assumed field line rotation and the current distribution in the 
magnetosphere.\cite{beskin09,okamoto97} In the low inertia (high magnetization) limit and flat Minkowski space, 
the GSE (without differential rotation) reduces to the original pulsar equation and describes the solution of the 
time-independent set of FFE equations. An example of a possible magnetic field structure emerging from the disk 
is shown in Fig.~\ref{fendt}.

\noindent Both, the FFE and the MHD fluid approach, essentially rely on the assumption that charge separation 
is/remains negligible everywhere such that $\vec{E} \cdot \vec{B} = 0$ is satisfied over the length scale of interest. 
This could be a reasonable assumption for e.g., field lines connecting to the disk and the surrounding plasma:\\ 
In the laboratory frame, the rotation of the magnetic field will induce an electric field $\vec{E} =-\vec{v}_{\rm rot}/c 
\times \vec{B}$. If not screened, this could in principle be tapped for the acceleration of particles. By Gauss' law, 
the induced electric field is supported by a local charge density $\rho_e = \nabla \cdot \vec{E}/4\pi$, corresponding 
to a particle number density (commonly referred to as the Goldreich-Julian [GJ] density\cite{goldreich69}) 
well inside $r_{\rm L}$ of
\beq\label{goldreich-julian}
  n_{\rm GJ} = \frac{\Omega B \cos\theta}{2\pi e c} = 0.1~ \eta^{-1} \left(\frac{B}{10^4~\mathrm{G}}\right)
                                                              \left(\frac{10^8 M_{\odot}}{M}\right)~\cos\theta~[\mathrm{particles~cm}^{-3}]
\eeq where $\Omega = c/r_{\rm L}$, $r_{\rm L}$ is the typical light cylinder scale with $r_{\rm L}=\eta r_s$, and $\eta 
\sim (1-10)$ for disk and $\eta = 1/a \sim 1$ for (extreme) Kerr black hole magnetospheres. Here, $\theta$ denotes 
the field inclination. If the environment is plasma-rich ($n >  n_{\rm GJ}$), the ability of charges to move freely along 
the field lines will ensure (at least inside $r_{\rm L}$) that $\vec{E} \cdot \vec{B} \rightarrow 0$, i.e., the parallel electric 
field component essentially becomes screened. (As a natural consequence, gap-type particle acceleration in the 
induced electric field will be suppressed.) Applicability of the MHD approximation thus basically requires that the 
charge density is always larger than the critical Goldreich-Julian density. For self-consistency one may thus always 
wish to verify that this is satisfied within a numerically calculated flow.\cite{fendt04} The inner regions 
of AGNs are usually sufficiently plasma-rich to make this a useful working assumption for disk-driven outflows. If 
one conservatively assumes, for example, that accretion occurs close to the free fall timescale, i.e., $t_a\sim r/v_r$, 
$v_r  \sim \alpha v_f $, $v_f=(GM/r)^{1/2}$, with $\alpha$ the viscosity parameter, the characteristic particle 
number density $n \sim \dot{M} t_a/\pi r^3 m_p$ close to $r_s$ becomes    
\beq\label{adaf-density}
  n \sim 10^{12} \alpha^{-1} \left(\frac{\dot{M}}{\dot{M}_{\rm Edd}}\right)
                                                 \left(\frac{10^8 M_{\odot}}{M}\right)
                                                 \left(\frac{r_s}{r}\right)^{3/2}~[\mathrm{particles~cm}^{-3}]\,,
\eeq i.e., orders of magnitude higher than the minimum GJ density required to screen the fields. The physical 
situation is less obvious for black hole-driven jets. Here, the nature of the plasma source that replenishes 
charges which escape the system along open magnetic field lines (both, outgoing to infinity and ingoing across 
the horizon) is less well understood, and charge-starved funnel regions or gaps can occur for which the ideal 
MHD description may not be adequate.\cite{kirk11,lev11} 
Plasma injection is then most likely be related to electron-positron pair production (cross-field diffusion being 
negligible), either due to direct $\gamma\gamma$-interactions of MeV photons in a hot accretion flow and/or 
initiated by, e.g., the Compton up-scattering of ambient photons by electrons/positrons accelerating in a vacuum 
gap (e.g.,refs.\cite{blandford77,rees84,beskin92,punsly98,vincent10,mosci11,lev11}). For a number of conditions, 
this is believed to ensure a plentiful supply of charges, cf. also eq.~(\ref{paircreation}), so that the plasma around 
(e.g., above the gap) a "typical" AGN is commonly also considered to satisfy quasi-neutrality ($n^++n^- \gg 
n^- - n^+$).\cite{begelman84}\\

\begin{figure}[t]
\centering
\psfig{figure=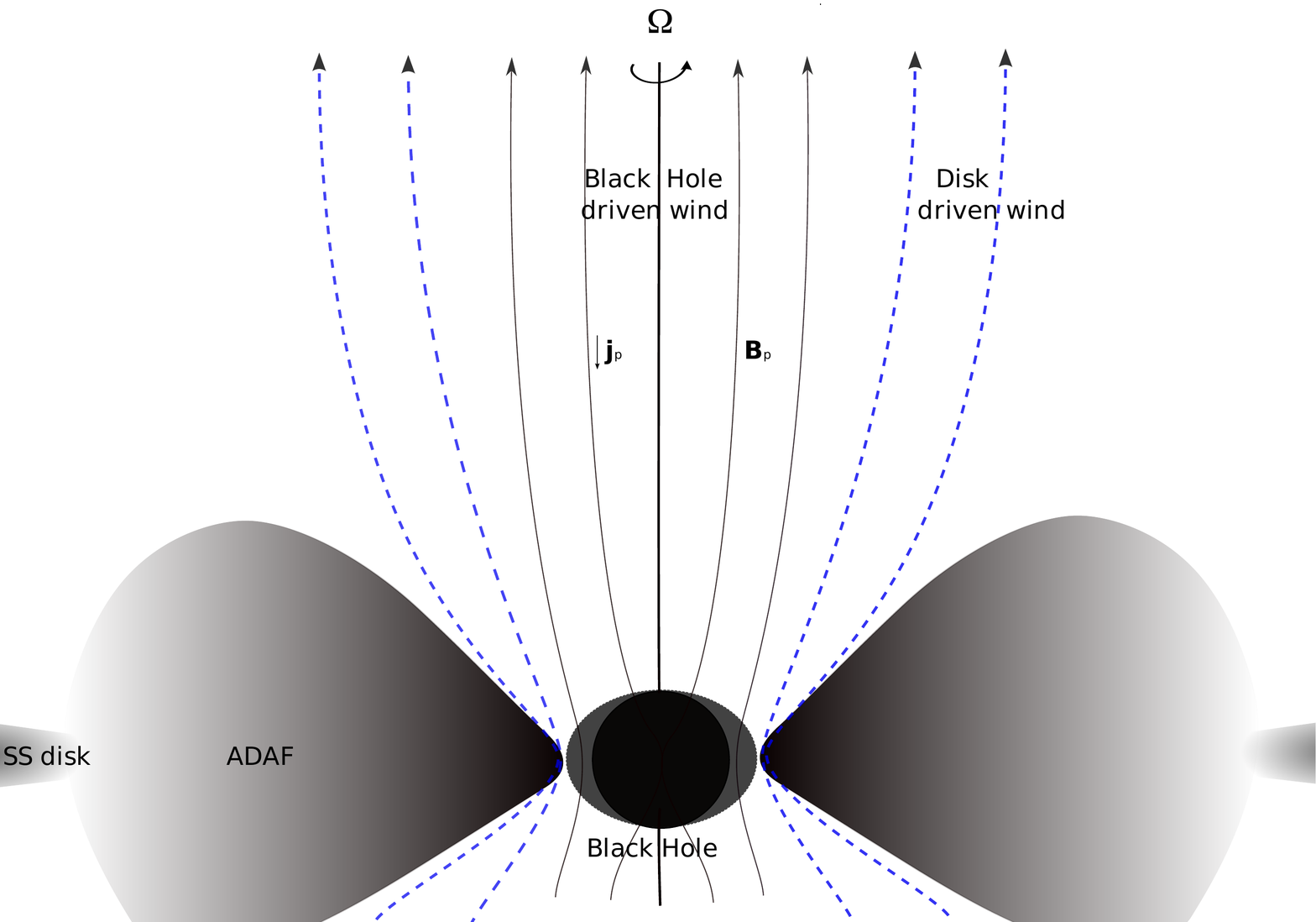, width=12truecm}
\caption{Schematic drawing of a possible field structure around a rotating black hole, with a black hole-driven 
wind surrounded by an outflow from the inner parts of the disk. The accretion flow may switch its configuration 
to a standard disk at some transition radius $r_t$. }\label{jet-drawing}
\end{figure}

\noindent Global 3d-relativistic MHD simulations of black hole accretion disks and magnetospheres have in 
recent years benefited from a proliferation of numerical approaches.\footnote{The interested reader is referred 
to the relevant reviews by, e.g. Fragile\cite{fragile08} and Krolik \& Hawley\cite{krolik10} for more details and 
references.} 
The results are widely considered to answer, at least partially, some of the longstanding questions with 
respect to magnetically-dominated (Poynting-flux-dominated)\cite{mckinney06} and matter-dominated 
jets.\cite{hawley06} Accordingly, the simulations confirm the general picture of a two-component sheath 
jet configuration with a fast, magnetically-dominated inner jet being surrounded by a slower moving, 
matter-dominated jet. As it appears, relativistic magnetically-dominated jets may in fact need an external 
agency (e.g. a disk wind) to facilitate efficient collimation (cf. also refs.\cite{fendt06,komissarov07}).
There are indications that the jet formation process does not require ordered, large-scale magnetic fields to 
be fed in from larger disc scales,\cite{mckinney06} and that the resultant outward-directed electromagnetic 
energy flux is consistent with the flux being powered by the Blandford-Znajek process.\cite{mckinney04} 
The first point, however, may need some further confirmation as other simulations still seem to need 
large-scale seed fields (e.g., ref.\cite{porth10}).\\
For the obtained matter-dominated jets, on the other hand, the main driving force seems to be a high-pressure 
corona (no evidence for the operation of a Blandford \& Payne-type process has been found yet), where strong 
(gas plus magnetic) pressure gradients ensure that matter is pushed into the jet (leading to extended mass 
loading) and accelerated along it.\cite{hawley06} This comes along with relatively slow jet speeds ($\Gamma_j 
\sim 1.5$) and large opening angles.

\noindent While very instructive, these global simulation results still face (apart from the usual methodological 
constraints, see e.g., ref.\cite{parks04}) a number of limitations whose influences need to be carefully assessed 
in order to allow detailed contact with observations. Among others, this relates to the facts that: (i) Not all of the 
global numerical schemes employed are fully conservative (e.g., in Hawley et al. only the internal, but not 
the total energy is explicitly conserved). (ii) Radiation fields, although observationally important, are as yet not 
fully included (e.g., in McKinney). This could obviously modify the flow dynamics and considerably influence 
the heating and cooling of the plasma. In fact, it has been argued that for powerful quasar-type sources, 
radiative driving may play an important role in shaping the outflow.\cite{kurosawa09} 
(iii) Usually, the plasma content is assumed to be conserved, which may not be appropriate in the vicinity
of the black hole, where pair creation is occurring all the time (e.g., refs.\cite{blandford02,lev11}).
(iv) The results can depend strongly on the chosen initial and boundary conditions. Since magnetic fields 
are, e.g., divergence-free, they seem never able to fully forget their (artificial) initial configurations, making 
the resultant jets very sensitive to these (observationally poorly known) conditions.\cite{beckwith08} 
(v) While global approaches rely on the ideal (infinite conductivity/zero resistivity) MHD limit, longterm 
simulation results seem to provide evidence for (artificial) reconnection-driven magnetic island formation. 
The occurrence of magnetic reconnection could significantly change the global field configuration and the 
plasma dynamics\cite{koide10} (for an illustration of the role of finite resistivity in the Newtonian limit, 
see ref.\cite{fendt09}; for the special relativistic case see ref.\cite{palenzuela09}). It is not clear to which 
extent this might invalidate some of the numerical results obtained so far with ideal MHD.

Nevertheless, for the purpose of our investigation here, we may visualize a simplified field structure as 
shown in Fig.~\ref{jet-drawing}.

\section{PARTICLE ACCELERATION IN BLACK HOLE-JET MAGNETOSPHERES}

\subsection{Direct electric field acceleration and the Blandford-Znajek process}
A black hole embedded in an external poloidal field $B_p$ and rotating with angular velocity $\Omega_H=
a (c/2 r_H)$, Eq.~(\ref{Omega_H}), will induce an electric field of magnitude $|\vec{E}| \sim (\Omega_H r_H) 
B_p/c$ corresponding to a permanent voltage drop across the horizon of magnitude $\Phi \sim r_H |\vec{E}| 
=(a/2) r_H B_p$. In terms of the electric circuit analogy, a black hole can act like a unipolar inductor (battery) 
with non-zero resistance, so that power can be extracted by currents flowing between its equator and poles. 
For reasonable astrophysical parameters, the potential drop (noting that one statvolt [cgs] corresponds to 300 
V [SI]) is of order (cf. ref.\cite{thorne86})
\beq\label{potential} 
 \Phi \sim 2 \times 10^{19} ~ a~(1+\sqrt{1-a^2}) \left(\frac{M}{10^8 M_{\odot}}\right) 
                                                   \left(\frac{B_p}{10^4~\mathrm{G}}\right)~\mathrm {[V]} 
\eeq If a charged particle can fully tap this potential, acceleration up to ultra-high energies of $E = Z e \Phi 
\sim 3 \times 10^{19} Z (M/10^8 M_{\odot}) (B_p/10^4~\mathrm{G})$ eV may become possible. Rapidly 
spinning and massive black holes have thus often been considered as astrophysical candidates for the
energization of UHE cosmic rays (cf. \S~\ref{UHE_sources}). As noted above, however, the charge density 
in the vicinity of accreting black holes may well be so high that a significant fraction of this potential is 
screened and thus no longer available for particle acceleration (see, however, also ref.\cite{neronov09}). 
It seems thus more appropriate to define an effective potential where the available length scale (gap 
height $h$) is explicitly taken into account, i.e., 
\beq\label{gap_potential}
 \Phi_e \sim \Omega_H  \left(\frac{h}{r_H}\right)^2 r_H^2 B_p/c = \left(\frac{h}{r_H}\right)^2 \Phi\,. 
\eeq Accordingly, the characteristic rate of energy gain can be expressed as 
\beq\label{BHaccel}
    \frac{dE}{dt} = Z e \Phi_e c/h\,,
\eeq with corresponding acceleration timescale $t_{\rm acc} = \gamma m_0 c h/(Z e \Phi_e)$.\\
Equation~(\ref{potential}) yields a strong electromotive force (EMF) which in the Blandford-Znajek (BZ)
scenario is assumed to drive electric currents along the fields and to generate a power output $P_{\rm BZ}=
\Phi I =\Phi^2/(\Delta Z_s)$. As the event horizon does not have perfect conductivity but instead acts like free 
space with respect to the propagation of electromagnetic waves, its surface resistivity $Z_s$ corresponds to 
the impedance of free space (377 Ohm = $4\pi/c$ [stat-ohm], with c in cgs).\cite{damour78,thorne86} 
Thus, if current closure can be achieved in the black hole environment (for field lines emerging between 
pole and equator, i.e. with resistance $\Delta Z_s \sim Z_s/4 \sim 100$ Ohm), a considerable electromagnetic 
(Poynting-flux-driven) power output of the order
\beq\label{pmax}
 P_{\rm BZ} \simeq 5 \times 10^{43} a^2 (1+\sqrt{1-a^2})^2 k \left(\frac{M}{10^8 M_{\odot}}\right)^2 
                                     \left(\frac{B_p}{10^4~\mathrm{G}}\right)^2~\mathrm{erg/s}\,.
\eeq might be generated, where $a$ is the dimensionless spin parameter and $k \sim 1$ is a numerical factor 
accounting for the field geometry. To support such an electromagnetic power output, however, an electric current 
$I = (P_{\rm BZ}/\Delta Z_s)^{1/2} \sim 7 \times 10^{26} \mathrm{statampere} \simeq 2 \times 10^{17} \mathrm{A}$ 
must flow within the magnetosphere, requiring a minimum steady pair injection (creation) rate of $dN_e/dt = I/Q \sim 
10^{36}$ particles/s. The implied (minimum) particle number density would then be close to the Goldreich-Julian 
density estimated in eq.~(\ref{goldreich-julian}). Even if one neglects cross-field diffusion, this density could, as 
shown later on, most likely be maintained by pair creation alone, cf. eq.~(\ref{paircreation}).\\
The above analysis, when put on a more rigorous basis (cf. the "Membrane paradigm"\cite{thorne86}), 
reveals a number of subtleties that have been the center of much attention and debate. To mention a few:\\
(1) In reality, the BZ power depends on the assumed magnetic field topology. The angular velocity of the field
lines $\Omega_f$, for example, does not simply coincide with the angular velocity of the black hole $\Omega_H$, 
but is instead dependent on the global field structure. Even if the impedance-matched case (where the resistance
of the BH "battery" equals the one of the "load" at large distances, i.e., where $\Omega_f=\Omega_H/2$) becomes 
established with time such that maximum power extraction is ensured,\cite{komissarov01} the numerical factor 
$k$ might well be smaller than one (e.g., $k \leq 0.2$ for a homogeneous magnetic field, see ref.\cite{beskin00}). 
In general, for an Eddington-limited system, the BZ process cannot extract energy at a rate much larger than 
$0.29~L_{\rm Edd}/\eta$, where $\eta\sim 0.1$ in the canonical conversion efficiency (cf. eq.~[\ref{maxLrot}]; 
also refs.\cite{menon05,dermer08}). Recent studies of powerful FR~II radio galaxies suggest that these sources can 
have jet powers comparable to the accretion power $P \sim \dot{M} c^2$ (e.g., ref.\cite{fernandes11}). This is much 
higher than expected from current, global relativistic MHD simulations of the BZ-process, that start off from a 
no-net-magnetic flux configuration, and may indicate that large-scale magnetic flux needs to be accreted in 
order for the process to operate efficiently enough (see ref.\cite{punsly11} for details).\\
(2) The "membrane" analogy, according to which the black hole horizon can be regarded as unipolar inductor 
with non-zero resistivity, has been fundamentally criticized on physical grounds\cite{punsly90} (cf. 
also ref.\cite{levinson04} for a response). As the horizon is causally disconnected from the external space and 
the outgoing flow, it actually cannot act as a unipolar inductor, which has led to general doubts concerning 
the nature of the BZ process. There are at least two (related) responses to this problem: (i) First, within the 
Grad-Shafranov MHD approach it can be shown that the "boundary condition" determining the potential 
energy loss of a rotating black hole is in fact dictated by the physical parameters in the pair creation region 
outside and not at the horizon.\cite{beskin00,okamoto06} This gives about the same power output and 
illustrates that the nature of the BZ process does not have to rely on a questionable causal connection 
between the event horizon and the outer magnetosphere. 
(ii) Secondly, the physical source of the BZ process does not -- when correctly interpreted -- lie with the horizon 
itself, but with the ergosphere. As plasma is unavoidable forced into co-rotation within the ergosphere, the 
magnetic field (assumed to be frozen in the plasma) is forced into co-rotation as well. While the horizon is 
indeed causally disconnected from the outgoing wind, the ergosphere is not and the BZ mechanism can still
be recovered.\cite{komissarov04,levinson06a}\\ 
(3) In the BZ scenario, black holes are assumed to be surrounded by plenty of charges, ensuring that an 
electric current can flow along the poloidal magnetic fields. This poloidal current then results from the 
screening of the vacuum electric field. On the other hand, to sustain a stationary current, the electric field 
should retain small unscreened components and not become completely screened by a redistribution of 
electric charges. Otherwise, the magnetospheres would no longer be able to drive electric currents (cf. 
ref.\cite{punsly90}). As it turns out, however, such a final state appears not possible within the ergosphere of 
a rotating black hole:\cite{komissarov04} 
The electric field, which is induced "gravitationally", cannot become completely screened by any static 
distribution of electric charges, i.e., rotating black holes do not allow stationary solutions with completely 
screened electric fields and vanishing poloidal currents.\\ 
(4) The BZ process of extracting the rotational energy of black holes appears to be very similar, but not 
simply equivalent to the mechanical Penrose process.\cite{penrose69} Because plasma entering the 
ergosphere is forced to co-rotate with the black hole, magnetic field lines penetrating the ergosphere will 
get twisted by the differential rotation of the plasma. This twist of magnetic field lines propagates outwards 
along the field lines (against the infalling plasma flow) as a torsional Alfv{\'e}n wave, establishing an 
outgoing electromagnetic energy flux. The magnetic tension of the bent field lines, on the other hand, 
decelerates the plasma in the ergosphere, imparting opposite angular momentum. The ergospheric 
plasma thus moves to lower energy orbits and the total (hydro- and electrodynamic) energy-at-infinity 
quickly decreases to negative values.\cite{koide03} While it seems that initially the resultant Penrose 
energy extraction process essentially operates via negative hydrodynamic energy-at-infinity, longterm 
simulations suggest that regions of negative hydrodynamic energy-at-infinity eventually disappear, with 
the electromagnetic BZ process still continuing.\cite{komissarov05}  
A more general analysis suggests, however, that the condition, for which the direction of the Poynting 
energy flux (at infinity) is opposite to that of the (local) field flow, is not simply equivalent to the Penrose 
one for negative energy-at-infinity (see, e.g.~\cite{komissarov08} for more details).\\
(5) According to recent results based on general relativistic MHD simulations, the maximum BZ jet 
power appears to be sensitive to the assumed jet-disk geometry. In the presence of, e.g., a thick disk 
(with effective thickness $(H/R) \sim 1$) the jet power may actually scale more strongly with black hole 
spin, i.e., $P_{\rm BZ} \propto a^4$ instead of $P_{\rm BZ}\propto a^2$ for a thin disk.\cite{tchek10} 
If this is confirmed by further research, it may offer important clues for understanding the 
radio-loud/radio-quiet dichotomy in AGN.

\subsection{Direct electric field acceleration close to the light cylinder}\label{beskin}
In early applications of the BZ process, charged particles (electric currents) are considered being able to 
cross the field lines only in the "load" region far away from the black hole.\cite{macdonald82}  
Accordingly one may expect significant particle acceleration to occur mainly within that region. However, this
needs not necessarily to be the case as has been demonstrated for pulsar magnetospheres:\cite{beskin_rafi00} 
Within ideal MHD, the structure of the magnetosphere is essentially determined by the supposed field line 
rotation and the assumed electric current distribution (cf. \S~\ref{magnetosphere}). If the longitudinal electric 
current would become sufficiently small, the freezing-in condition will break down so that electrons, deviating 
from the field line, may become able to sample the full induced electric field and thereby get efficiently accelerated 
along the z-axis close to the black hole.\cite{beskin83} This would obviously allow acceleration beyond the 
asymptotic flow speed in ideal MHD $\gamma \sim \sigma_m^{1/3}$ (e.g., ref.\cite{michel91}). As shown by Beskin 
\& Rafikov\cite{beskin_rafi00} for a quasi-monopole magnetic field configuration, this could happen in a narrow 
boundary layer of thickness $\Delta r $ close to the generalized light surface 
\beq\label{light-surface}
r_{ls} = \frac{r_{\rm L}}{(2\eta_c)^{1/4}}\,,
\eeq where $|\vec{E}| = |\frac{\vec{v}_{\rm rot}}{c} \times \vec{B}| \simeq |\vec{B}|$. Here, $\eta_c >0$ 
characterizes the strength of the longitudinal current in terms of the Goldreich-Julian current, $j_{\|} = 
(1-\eta_c) j_{\rm GJ}$. For small longitudinal currents, i.e. $\eta_c \rightarrow 1$, the light surface coincides 
with the classical light cylinder 
\beq
r_{ls} \rightarrow r_{\rm L}=\frac{c}{\Omega}\,. 
\eeq Sampling an electric field of strength $\sim |\vec{B}|$, electron acceleration (in the absence of losses)
is then approximately described by 
\beq
\frac{d\gamma}{dr} \simeq \frac{e B}{m_e c^2}\,
\eeq The maximum electron Lorentz factor that can be achieved in the boundary layer of width $\Delta r$  
thus becomes
\beq
 \gamma \sim \frac{d\gamma}{dr} ~ \Delta r\,.
\eeq As the thickness of the layer depends on the charge density, i.e., 
\beq
\Delta r \sim \frac{r_{\rm L}}{\lambda}\,, 
\eeq where $\lambda = n_e/n_{\rm GJ} \gg 1$ is the multiplicity factor and $n_{\rm GJ} = \Omega B/(2\pi c e)$ 
the Goldreich-Julian density,\cite{beskin_rafi00}  the maximum Lorentz factor becomes
\beq
 \gamma \sim \sigma_m
\eeq where $\sigma_m = B^2/(8\pi n_e m_e c^2)$ is the Michel\cite{michel69} magnetization parameter. 
Thus, in a narrow boundary layer close to the light cylinder $r_{\rm L}$ almost all of the electromagnetic 
energy may be transformed into the kinetic energy of particles.

\subsection{Centrifugal acceleration close to the light cylinder}
A somewhat different approach, that does not rely on direct electric field acceleration but takes inertial 
(centrifugal) and radiation reaction effects explicitly into account, has been explored in the literature as 
well.\cite{machabeli94,machabeli96,gangadhara97,rieger00,xu02,osmanov07,rieger08a,rieger08b}\\
This approach assumes that the plasma density is in fact high enough to ensure that the parallel electric field 
component is effectively screened out up to distances (of at least) very close to the light cylinder, and that the 
magnetic field structure is such as to allow a significant fraction of the electromagnetic energy to be transformed 
into the kinetic energy of particles through magneto-centrifugal effects close to $r_{\rm L}$. This goes along with 
the requirement that the longitudinal electric current is small enough.
Consider for illustration a charged test particle, co-rotating with the magnetic field and gaining energy while moving 
outwards. Under most circumstances, an energetic electron will quickly lose its perpendicular energy component 
due to strong synchrotron losses and may thus be considered as just sliding along the field line. The motion of such 
a particle along a field line may be conveniently analyzed in the framework of Hamiltonian dynamics: The Lagrangian 
$L$ for a particle with rest mass $m_0$ moving along a (two-dimensional) relativistically rotating field line (angular 
velocity $\Omega=c/r_{\rm L} =$ const.) is simply given by
\beq
L = -m_0 c^2 (1-v_r^2/c^2-v_{\phi}^2/c^2)^{1/2}.
\eeq where $v_r = \dot{r}$ and $v_{\phi}= \Omega r + \dot{r} \frac{B_{\phi}}{B_r}$ (note, that as the field is 
considered to be swept back, $B_{\phi}/B_r < 0$). As $L$ is not explicitly time-dependent, the Hamiltonian $H=
\dot{r} P - L$, with $P=\partial L/\partial \dot{r}$ the generalized momentum, is a constant of motion (Noether's
theorem). Using the above relations, one easily finds (cf. also ref.\cite{contopoulos99})
\beq\label{ham}
H = \gamma m_0 c^2 \left(1-\frac{\Omega r}{c^2} v_{\phi}\right) = \mathrm{const.}\,,
\eeq where $\gamma=(1-v_r^2/c^2-v_{\phi}^2/c^2)^{-1/2}$ is the Lorentz factor of the particle. This can be
generalized to the case where $\vec{B}$ and $\vec{\Omega}$ are inclined at an angle $\theta$ by replacing
$r \rightarrow (r \sin\theta)$ on the rhs of eq.~(\ref{ham}) and, correspondingly, in Ferraro's law of isorotation. 
Note that the above expression might have also been obtained from the standard Bernoulli equation (energy 
flow conservation). Provided the product $\dot{r} B_{\phi}/B_r$ remains sufficiently small, this implies 
\beq
\gamma(r) \propto \frac{1}{(1-r^2/r_{\rm L}^2)}\,.
\eeq Thus, for a particle approaching the light cylinder $r \rightarrow r_{\rm L}$, the Lorentz factor may increase 
dramatically as long as co-rotation holds. However, even in the single particle approach the Lorentz factor cannot 
become arbitrarily large, as from a formal point of view the validity of the approach demands that at least $\omega_c 
\leq \Omega$ (see also the breakdown constraint below),\cite{istomin09} where $\omega_c = eB/(\gamma m_0 
c)$ is the relativistic gyro-frequency. Using Eq.~(\ref{ham}) and the definition of $\gamma$, the characteristic 
acceleration time scale can be expressed as\cite{rieger08a} 
\beq\label{t_centri}
      t_{\rm acc}=\frac{\gamma}{\dot{\gamma}} \simeq \frac{1}{2\,\Omega\, \tilde{m}^{1/4}\gamma^{1/2}}\,,
\eeq valid for $\gamma \gg 1$, where $\tilde{m}=1/(\gamma_0^2\,[1-r_0^2/r_{\rm L}^2]^2)$ depends on the 
initial injection conditions. Under more realistic circumstances, achievable particle Lorentz factors are expected 
to be limited either (i) by radiative losses (curvature or Compton), (ii) by the breakdown of the (single particle)
bead-on the wire approximation (relevant for, e.g., protons), or (iii)  - if we consider an ensemble of particles -  
by the particles' inertia overcoming the tension in the field lines (i.e., breakdown of plasma co-rotation, e.g., 
see ref.\cite{osmanov_r09}). In the simplest case the breakdown (ii) corresponds to the situation where the Coriolis 
force becomes comparable to the Lorentz force. It is tantamount to requiring that the inverse of the relativistic 
gyro-frequency $\omega_c$ remains smaller than the acceleration timescale.\cite{rieger08b} This translates 
into an upper limit on achievable Lorentz factors of
\beq\label{max-centrifugal-proton}
 \gamma_{\rm max} \simeq  2 \times 10^8~\tilde{m}^{-1/6} \left(\frac{B(r_{\rm L})}{100~\mathrm{G}}\right)^{2/3}
                                           \left(\frac{m_e}{m_0}\right)^{2/3}\left(\frac{r_{\rm L}}{10^{14}\mathrm{cm}}\right)^{2/3}
\eeq where $B(r_{\rm L})$ denotes the field strength at the light cylinder $r_{\rm L}$.

\subsection{Magnetic reconnection at the jet base}
Efficient magnetic reconnection (annihilation of magnetic fields) could possibly take place close to the black hole 
if there are suitable field regions with opposite polarities. In the reconnection process, the energy stored in these 
fields is released as kinetic energy and heat (increase in entropy) (cf. ref.\cite{zweibel09} for review). This requires 
finite resistivity (i.e., breakdown of ideal MHD and the frozen-in condition) and becomes important in regions with 
large magnetic field gradients (for application to AGN see, e.g., refs.\cite{lesch91,kirk04}). By Ampere's law, $\nabla 
\times \vec{B}= (4\pi/c)\vec{j}$, these large field gradients are associated with large current densities $\vec{j}$ 
("current sheets"). Thus, even for a relatively high electrical conductivity $\sigma_e$, significant ohmic losses 
($j^2/\sigma_e$) might occur. 

\begin{figure}[ht]
\centering
{\psfig{figure=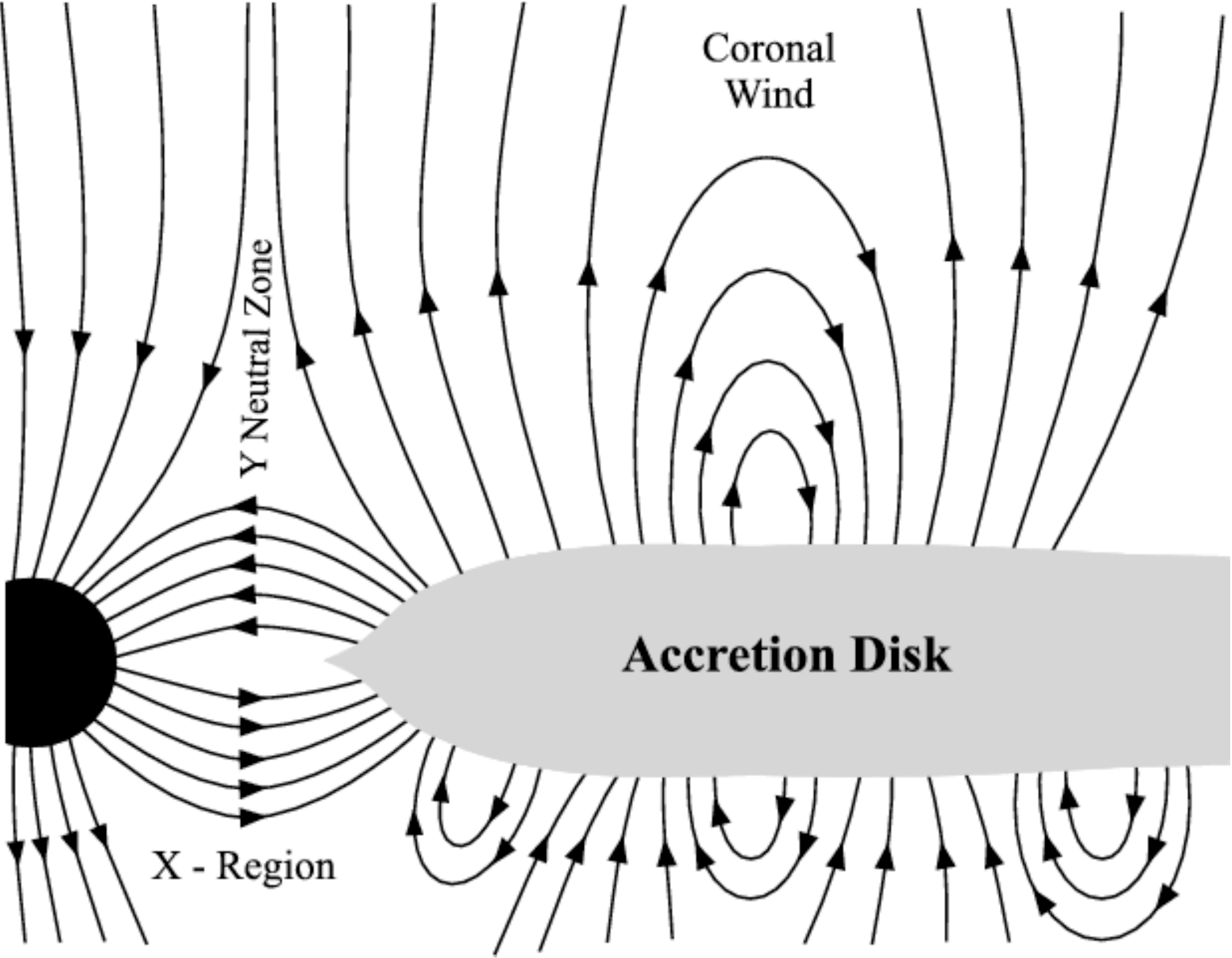, width=6.5truecm}}{\psfig{figure=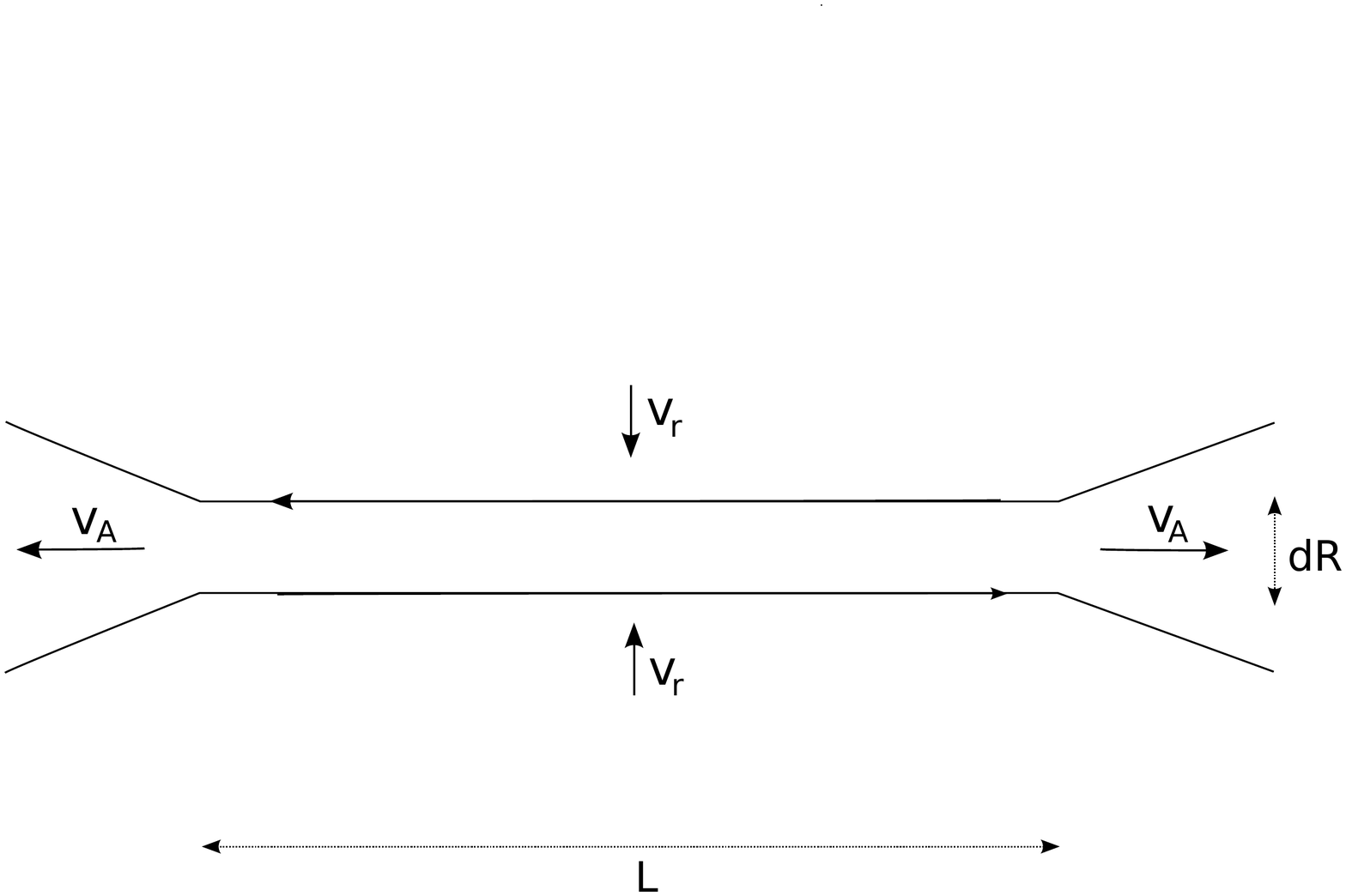, width=6.5truecm}}
\caption{Left: Schematic drawing of the assumed magnetic field geometry. Acceleration could occur in the 
magnetic reconnection region at the Y-type neutral zone. From de Gouveia Dal Pino and Lazarian\protect\cite{dalpino05}. 
Right: Illustration of a simple reconnection region. Two oppositely directed magnetic fields in a plasma are 
carried towards the neutral line at speed $v_r$ over a characteristic length scale $L$. There is a layer of 
width $dR$ in which the field reconnects. Reconnected field and plasma are expelled at speed $v_A$.}
\label{dal_pino}
\end{figure}

\noindent The rate of magnetic reconnection is controlled by the geometry of the dissipation region. In the 
simplest (Sweet-Parker) picture, see Fig.~\ref{dal_pino}, the inflow of plasma and magnetic field into the 
reconnection region at speed $v_r$ is balanced by an outflow at speed $v_A$. The characteristic outflow 
speed for a magnetized plasma is the Alfv\'{e}n speed $v_A=B/(4\pi n_c m_p)^{1/2}$. Mass conservation 
approximately implies $L~v_r \sim dR~v_A$, i.e., $v_r \sim (dR/L) v_A$. From the (steady state) induction e
quation, the width can be estimated to be $dR \sim \lambda_m/v_r$, where $\lambda_m=c^2/(4\pi \sigma_e)
\propto 1/R_m$ is the magnetic diffusivity ($R_m \sim v_A L/\lambda_m \gg 1$ the magnetic Reynolds number), 
measuring the ability of a magnetic field to diffuse through a plasma. This combines to $v_r \sim v_A/R_m^{1/2} 
\ll v_A$, so that the reconnection process would only proceed at very slow speeds, the classical problem 
associated with (Sweet-Parker-type) reconnection. 

\noindent However, plasma micro-turbulence or wave-particle interactions might lead to an enhanced 
("anomalous") resistivity (increasing $\lambda$), allowing faster reconnection rates with $v_r  \rightarrow 
v_A$. Such a situation has been envisaged to take place at the interface between a black hole magnetosphere 
and a coronal wind, where poloidal magnetic field reversal may occur, see Fig.~\ref{dal_pino}.\cite{dalpino05,dalpino10} 
The maximum rate of magnetic energy that can be extracted from the reconnection zone in the corona (above 
and below the disk) would then be on the order of
\beq
  P_B \sim (B^2/8\pi) v_r (4\pi R) L \sim (B^2/8\pi) v_A (4\pi R) dR\,,
\eeq where $B$ is the magnetic field at the reconnection zone, $R \sim R_s$ is of the order of the inner radius 
of the disk, and $v_A \sim c$ for a highly magnetized plasma. In analogy to Ampere's law, we can roughly 
estimate the width $dR$ of the current sheet for which the resistivity must be anomalous\cite{dalpino05}
\beq\label{rc_width}
 dR \sim \frac{c \Delta B}{4\pi n_c Z e v_{\rm th,c}} \sim 200~\frac{m_p c^2}{Z e B}\,,
\eeq where $v_{\rm th,c}\simeq (kT/m_p)^{1/2}\sim c/100$ is the typical thermal velocity of the ions in the X-ray 
($kT \sim 100$ keV) emitting corona, and where $\Delta B \sim 2 B \sim 10^5~(10^8 M_{\odot}/M)^{1/2}$ G (cf. 
\S~\ref{mfields}) denotes the change of the magnetic field across the reconnection region. This gives a relatively
modest power of
\beq
 P_B \sim 5 \times 10^{37} \left(\frac{M}{10^8 M_{\odot}}\right)^{1/2}~\mathrm{erg/s}\,.
\eeq Yet, while not sufficient to account for radio-loud AGNs, this power could perhaps be sufficient to explain 
the observed radio luminosities of radio-quiet AGNs (see Fig. 3 in ref.\cite{dalpino10}).\\ 
During the reconnection process, efficient acceleration of supra-thermal particles in the inductive electric field 
$\vec{E}_r = \vec{v}_r \times \vec{B}/c$ perpendicular to the (two-dimensional) reconnection fields can occur 
as long as their Larmor radii do not exceed the width of the current sheet.\cite{lesch91} Ignoring energy losses, 
this would suggest that in the considered scenario (cf. eq.~[\ref{rc_width}]) relativistic energies up to $\gamma_p 
\sim 2 \times 10^2$ (protons) and $\gamma_e \sim 4 \times 10^5$ (electrons) might become achievable (cf. also
ref.\cite{dalpino05} for possibility of further acceleration).\\ 
The details of the motion of charged particles in current sheets are non-trivial,\cite{craig02} and obtained 
results are dependent on the (time-dependent) sheet structure and, e.g., the strength of the linking 
magnetic field perpendicular to the sheet. Possible particle spectra discussed in the literature for fast 
magnetic reconnection include quasi-monoenergetic distributions (e.g., ref.\cite{lesch92}) and power-law 
particle spectra $n(\gamma) \propto \gamma^{-s}$ with $s \simeq 1$ (e.g., ref.\cite{zenitani01}) or $s \simeq 
1.5$ (ref.\cite{romanova92}).

\section{RADIATIVE PROCESSES}
The interactions of energetic particles with (photon or magnetic) fields and/or ambient matter usually provide
important efficiency constraints for the above mentioned acceleration mechanisms. Some of these loss 
processes and their possible relevance are discussed below.

\subsection{Synchrotron Radiation}
A charged particle with Lorentz factor $\gamma$ and charge number $Z$ spiraling in a magnetic field $\vec{B}$ 
emits synchrotron radiation with a mean frequency
\beq
  \left< \nu \right> =0.31~\nu_{cs}
\eeq where $\nu_{cs}=(3/4\pi) \Omega_e\gamma^2 \sin\theta$, $\Omega_e = Z e B/(m_0 c)$ is the gyro-frequency
and $\theta$ the pitch angle between the magnetic field and the velocity vector. The energy loss rate, $-dE/dt$, or 
the power emitted by a single particle is 
\beq
  P_{cs} = \frac{2}{3} \frac{Z^4 e^4}{m_0^2 c^3} \gamma^2 B^2 \sin^2\theta\,, 
\eeq so that the characteristic cooling timescale $t_s = E/|dE/dt| = \gamma m_0 c^2/P_{cs}$ becomes
\beq\label{s_cool}
   t_s = \frac{3~m_0^3 c^5}{2~Z^4 e^4\gamma B^2 \sin^2 \theta} 
             \simeq 520 \left(\frac{m_0}{m_e}\right)^3 \left(\frac{10^3 \mathrm{G}}{B}\right)^2 
             \frac{1} {Z^4 \gamma \sin^2\theta} ~ \mathrm{sec}\,.
\eeq In terms of the mean frequency $\left<\nu \right>$, one obtains
\beq
   t_s \simeq 5.9 \times 10^{11} Z^{-7/2} (m_0/m_e)^{5/2}(B\sin\theta)^{-3/2} \left<\nu\right>^{-1/2}~\mathrm{sec}\,.
\eeq The characteristic cooling time of a proton, for example, emitting synchrotron radiation at $\left<\nu\right>=2.4 
\times 10^{26}$ Hz (energy of 1 TeV) is thus of order $t_s \simeq 64~(B \sin\theta)^{-3/2}$d.\\
The synchrotron spectrum (power per unit frequency) $I(\nu)$ emitted by a single particle is a power law $I(\nu) 
\propto \nu^{1/3}$ for $\nu \ll \nu_{cs}$, and decreases exponentially for $\nu \gg \nu_c$. For most practical purposes, 
it can be reasonably approximated by $I(\nu) \propto (\nu/\nu_{cs})^{0.3} \exp(-\nu/\nu_{cs})$.\\
Equation~(\ref{s_cool}) shows that for almost all pitch angles $\theta$, energetic electrons are expected to quickly 
radiate away their perpendicular momentum components in the strong magnetic fields close to the black hole. 
Electrons in the black hole magnetosphere may thus be regarded as mainly occupying the ground ($n=0$) Landau 
state, moving quasi-one-dimensionally along the field lines. In order to emit synchrotron radiation, particles in the 
ground Landau state would have to be excited to higher Landau levels (by acquiring non-vanishing pitch angles). 
This is not excluded but may become possible through, e.g., Compton scattering or non-resonant diffusion in pitch 
angles.\cite{machabeli00,osmanov10a,osmanov10b} 

\subsection{Curvature Radiation}
In the black hole magnetosphere, and in particular closer to the light surface, the curvature of the magnetic 
field may no longer be negligible, so that particles moving along the fields might efficiently lose energy due 
to curvature radiation. In analogy to synchrotron radiation, curvature radiation can be described as emission 
from relativistic charged particles moving around the arc of a circle chosen such that the actual acceleration 
corresponds to the centripetal one, i.e. with curvature radius $R_c = \gamma m_0 c^2/(ZeB\sin\theta)$ (e.g., 
ref.\cite{ochelkov80}). The critical frequency where most of the radiation is emitted is given by
\beq\label{curv_freq} 
\nu_c \simeq \frac{3 c}{4\pi R_c} \gamma^3\,,
\eeq which for a curvature radius of, e.g. $R_c = r_s =  3 \times 10^{13} (M/10^8M_{\odot})$ cm yields 
$\nu_c \simeq 2 \times 10^{26}~(\gamma/10^{10})^3$ Hz or a curvature photon energy of about $1~(\gamma/
10^{10})^3$ TeV. Like synchrotron emission, the spectrum $I(\nu)$ produced by curvature radiation of a single 
particle is a power law $I(\nu) \propto \nu^{1/3}$ for $\nu \ll \nu_c$, and decreases exponentially for $\nu \gg 
\nu_c$. The energy loss rate or total power radiated away by a single particle (charge number $Z$) is 
\beq\label{curv_power}
  P_c = \frac{2}{3} \frac{Z^2 e^2 c}{R_c^2} \gamma^4\,. 
\eeq 
The characteristic cooling timescale $t_c = \gamma m_0 c^2/P_c$ thus becomes
\beq
  t_c \simeq 180 \frac{R_c^2}{Z^2} \left(\frac{m_0}{m_e}\right) \frac{1}{\gamma^3}\,.
\eeq In the absence of other damping mechanisms, a balance with cooling $dE/dt = P_c$ (using 
eq.~[\ref{BHaccel}]) would imply that direct electric field acceleration may account for particle Lorentz 
factors up to (cf. ref.\cite{levinson00}) 
\beq\label{gap-maximum}
   \gamma_{c, \rm max} \simeq 10^{10} ~ \frac{a^{1/4}}{Z^{1/4}} \left(\frac{M}{10^8 M_{\odot}}\right)^{1/2} 
                                                            \left(\frac{B_p}{10^4 \mathrm{G}}\right)^{1/4}
                                                            \left(\frac{R_c}{r_H}\right)^{1/2}\left(\frac{h}{r_H}\right)^{1/4}\,,
\eeq provided the potential (gap height $h$, spin $a$, cf. eqs.~[\ref{potential},\ref{gap_potential}]) is such that 
these Lorentz factors can in principle be achieved. Note that eq.~(\ref{gap-maximum}) does not depend on 
the particle (e.g., electron or proton) mass. The maximum energy of the emitted curvature photons becomes
\beq\label{max_curv_photon}
\epsilon_c =\frac{3}{2} \frac{c\hbar\gamma_{c,\rm max}^3}{R_c} \simeq 2~a^{3/4}
                       \left(\frac{M}{10^8 M_{\odot}}\right)^{1/2} \left(\frac{B_p}{10^4 \mathrm{G}}\right)^{3/4}
                       \left(\frac{R_c}{r_H}\right)^{1/2}\left(\frac{h}{r_H}\right)^{3/4}\,\mathrm{TeV}\,.
\eeq 
Equation~(\ref{gap-maximum}) suggests that the maximum energy for a charged particle of mass 
$\mu m_p$ may exceed $\sim 10^{19} \mu$ eV only in massive black hole environments. The magnetic 
field strength typically scales as $B \propto M^{-1/2}$ (cf. \S~\ref{mfields}), so that efficient UHE proton 
acceleration would require a very massive source. Equation~(\ref{max_curv_photon}) then suggests 
that this should be accompanied by significant VHE $\gamma$-ray emission.

\noindent As an illustration, let us suppose that particles are efficiently accelerated in the magnetosphere 
of a massive black hole, with their maximum energies only limited by curvature losses. As noted above, 
this could lead to detectable emission in the TeV energy band. For the gap to exist, the charge density 
has to be smaller than the Goldreich-Julian value, cf. eq.~(\ref{goldreich-julian}), i.e., $n \leq n_{\rm GJ} 
= \Omega B/(2\pi ec)$ (provided this density can be supported by the fields). 
Assuming a quasi mono-energetic particle distribution, the total curvature output at energy $\epsilon_c$, 
eq.~(\ref{max_curv_photon}), is then of order $L_c \simeq n ~\Delta V~P_c$. If acceleration takes place 
in a gap (of thickness $h$, with associated volume element $\Delta V \sim \pi \eta r_g^2 h$, where $\eta 
\leq 1$ is a geometrical factor), we have at maximum $P_c=dE/dt=e\Phi_e c/h$ (cf. eq.~[\ref{BHaccel}]). 
This would imply a maximum VHE luminosity of the order of (e.g., \cite{lev11})
\beqn\label{curvature_output}
 L_c &\simeq& n~(\pi \eta r_g^2 h) \left(\frac{e\Phi_e c}{h}\right)\\ \nonumber
         &\simeq& 8 \times 10^{43}\eta \left(\frac{n}{n_{\rm GJ}}\right)\left(\frac{B}{10^4\mathrm{G}}\right)^2 
                                  \left(\frac{M}{10^8 M_{\odot}}\right)^2 \left(\frac{h}{r_g}\right)^2 \frac{\rm erg}{\rm s}\,.
\eeqn It is interesting to note that the ratio $L_c/P_{\rm BZ}$, where $P_{\rm BZ}$ denotes the maximum 
Blandford-Znajek power, eq.~(\ref{pmax}), scales as $(h/r_g)^2$ (see also ref.\cite{lev11}).

\subsection{Inverse Compton Scattering}
Compton scattering of photons by relativistic electrons (Lorentz factor $\gamma \gg 1$) is usually called inverse 
Compton (IC) scattering. The IC power for single scattering of an isotropic photon distribution with energy density 
$U_{\rm ph} \simeq h \int_0^{m_ec^2/\gamma h} d\nu~\nu n_{\rm ph}(\nu)$ [erg/cm$^3$] in the Thomson limit $h
\nu \ll m_e c^2/\gamma$ (i.e., for negligible energy transfer in the electron rest frame) is
\beq\label{compton_power}
 P_{\rm IC} = -\frac{dE}{dt} = \frac{4}{3} \sigma_T c \gamma^2 U_{\rm ph}\,,
\eeq which gives a characteristic electron cooling timescale of
\beq\label{IC_cooling}
 t_{\rm IC} \simeq 3.1 \times 10^7 \frac{1}{\gamma U_{\rm ph}}\;\mathrm{sec}\,.
\eeq The photon scattering rate, i.e., the total number of incident and up-scattered photons per unit time is $c
\sigma_T n_{\rm ph}$, where $n_{\rm ph}=U_{\rm ph}/h\left<\nu\right>$. From $P_c = c \sigma_T n_{\rm ph} h 
\left< \nu_1\right>$ one deduces that the mean scattered photon frequency is
\beq
  \left< \nu_1 \right> = \frac{4}{3} \gamma^2 \left< \nu \right>
\eeq where $\left< \nu \right>$ is the average incident photon frequency. From the kinematics, the maximum (peak)
photon energy after scattering is $\nu_1 =4 \gamma^2 \nu$ for radiation with incident frequency $\nu$. Taking the 
Thomson condition into account, we have $h \nu_1\sim \gamma m_e c^2$. If Compton recoil cannot be neglected 
(i.e., if the Thomson regime does not apply), the full Klein-Nishina cross-section must be used, resulting in a reduced 
power output. Based on QED considerations, the total cross-section is given by the Klein-Nishina (KN) formula, for 
which two useful limits are
\beqn
\sigma &\simeq& \sigma_T \quad \mathrm{for}\; h\nu_r \ll m_e c^2
                               \nonumber \\
             &\simeq& \frac{3}{8} \sigma_T \left(\frac{m_e c^2}{h\nu_r}\right) \left[\ln\left(\frac{2h\nu_r}{m_e c^2}\right) 
                                + \frac{1}{2} \right]   \quad \mathrm{for}\; h\nu_r \gg m_e c^2\,.
\eeqn where $\nu_r$ is measured in the rest frame of the electron. In the extreme KN regime, the energy loss rate 
for a single electron is found to be\cite{blumenthal70} 
\beq
 - \frac{dE}{dt} \simeq \frac{3}{8} \sigma_T m_e^2 c^5 h^{-1} \int_{m_e c^2/\gamma h}^{\infty} d\nu~ 
                        \frac{n_{\rm ph}(\nu)}{\nu} \left[\ln\left(\frac{4\gamma h \nu}{m_e c^2} -\frac{11}{6}\right)\right]\,,
\eeq and is thus only weakly (logarithmically) dependent on the electron Lorentz factor $\gamma$. However, in 
contrast to the Thomson case, the photon now carries away a sizable fraction of the electron energy, so that the 
electron loses its energy in discrete steps.\\  
The spectrum emerging from single inverse Compton scattering generally depends on both, the seed photon spectrum
and the relativistic electron distribution. If the range of the seed photon spectrum is very narrow compared to the electron 
distribution (assumed to obey a power-law $n_e(\gamma) \propto \gamma^{-s}$ between $\gamma_{\rm min}$ and 
$\gamma_{\rm max}$), for example, the output spectrum is a power-law $\nu^{-(s-1)/2}$ over a dynamical range of 
$(\gamma_{\rm max}/\gamma_{\rm min})^2$, with index determined by the electron power-law index. On the other hand, 
if the electron distribution is very narrow compared to the photon distribution, the output spectrum resembles the seed
photon spectrum shifted in energy by a factor $\sim \left<\gamma^2\right>$.\\
If electrons are rotationally accelerated in the magnetosphere of a black hole embedded in an ambient photon field, 
Compton losses may significantly reduce achievable particle energies. In the Thomson regime, for example, inverse 
Compton upscattering of accretion disk photons will lead to particle energy losses on a timescale $t_{\rm IC} \propto 
\gamma^{-1}$ that decreases faster than $t_{\rm acc} \propto \gamma^{-1/2}$, eq.~(\ref{t_centri}), and so introduces a 
natural limitation. Balancing acceleration by cooling, eq.~(\ref{IC_cooling}), implies a characteristic upper limit of (e.g., 
ref.\cite{osmanov07})  
\beq\label{compton}
   \gamma_{\rm max} \sim  3 \times 10^6\, \frac{\sqrt{\tilde{m}}}{U_{\rm ph}^2} 
                                                 \left(\frac{10^{15}\mathrm{cm}}{r_{\rm L}}\right)^2\,,
\eeq assuming co-rotation to hold for such a range of Lorentz factors. Hence, only for highly underluminous AGN 
sources with target photon fields $L_t \ll L_{\rm Edd}$ ($U_{\rm ph} \sim L_t/[4\pi r_{\rm L}^2 c]$), will centrifugal 
acceleration allow to accelerate electrons to Lorentz factors $\gamma \gg 100$.\cite{rieger00,xu02} 

\subsection{Photo-meson production}\label{photo-meson}
Photo-meson production can become an important channel by which the kinetic energy of protons is transformed 
into high energy gamma-rays, electrons and neutrinos.\cite{kelner08} The kinematic threshold for single pion 
production (p+$\gamma \rightarrow p + \pi^0, n+\pi^+$) is given by 
\beq\label{photo-meson-threshold}
 2 E_p \epsilon_{\gamma} (1-\cos\theta)= (2 m_{\pi} m_p + m_{\pi}^2) c^4 \simeq 2.8\times 10^{17}~(\mathrm{eV})^2\,
\eeq where $E_p$ is the proton energy (assumed to be relativistic), $\epsilon_{\gamma}=h\nu$ is the target photon 
energy and $m_{\pi}$ the pion rest mass. Thus, in the rest frame of the proton, the photon energy has to exceed 
$\epsilon_{\gamma}' =m_{\pi} c^2 (1+m_{\pi}/2 m_p) \simeq 140$ MeV for the process to become possible. Close to 
the energy threshold, the process proceeds through single-pion production, while at higher energies multi-pion 
production channels start to dominate. The cross-sections for these inelastic processes are known from particle 
acceleration experiments. In first-order approximation, they can be expressed as a sum of two step-functions with 
$\sigma_1 = 0.34$ mbarn ($=3.4 \times 10^{-28}$ cm$^2$) in the interval 200 MeV $\leq \epsilon_{\gamma}' \leq$ 
500 MeV and $\sigma_2 = 0.12$ mbarn for  $\epsilon_{\gamma}'\geq 500$ MeV, and related inelasticities $K_{p,1}
=0.2$ and $K_{p,2}=0.6$, respectively.\cite{atoyan03} For the characteristic attenuation length we have $\lambda 
\sim 1/(n_{\gamma} \sigma K_p)$, where $n_{\gamma}$ is the photon number density [particles/cm$^3$], 
so that the characteristic proton cooling time $t_{\rm c} = \lambda/c$ becomes
\beq\label{photomeson}
   t_{\rm c} \sim \frac{5\times 10^{17} \mathrm{sec}}{n_{\gamma}}\sim 3 \times 10^6 
                          \left(\frac{\epsilon_{\gamma}}{1~\mathrm{keV}}\right) \left(\frac{10^{42} \mathrm{erg/s}}{L}\right) 
                         \left(\frac{R}{10^{14} \mathrm{cm}}\right)^2 \;\mathrm{sec}\,,
\eeq where $n_{\gamma}\simeq L/(4\pi R^2 c \epsilon_{\gamma})$ has been expressed in terms of the source 
luminosity $L$. If the inner disk surrounding the black hole would be of the standard (geometrically-thin, optically-thick) 
type, then the thermal disk radiation field would be dominated by emission from regions close to $r_s$, with energy 
peaking at $\epsilon_{\gamma}\sim 50$ eV, cf. eq.~(\ref{ss_disk_frequency}), and thereby provide a suitable ambient 
target field for protons with Lorentz factors above $\gamma_{p,t} \sim 2 \times 10^6~(50~\mathrm{eV}/\epsilon_{\gamma})$. 
Neglecting curvature losses and field screening, gap-type particle acceleration of protons up to ultra-high energies 
$\sim 10^{20}$ eV, cf. eq.~(\ref{gap_potential}), would then only be possible if mass accretion occurs at low rates 
corresponding to disk luminosities $L_d \lppr  0.01 L_{\rm Edd}$ (cf. ref.\cite{begelman91}).

\subsection{Bethe-Heitler (proton-photon) pair production}
At energies below the threshold for photo-meson production, the main channel for inelastic interactions of 
energetic protons with ambient photons is the direct production of electron-positron pairs, i.e. the process 
$p+\gamma \rightarrow e^+ + e^- + p$, e.g.refs.\cite{kelner08,mastichiadis05} The minimum energy (threshold 
condition) required for this process to become possible is 
\beq
  \gamma_p \epsilon_{\gamma} (1- \beta_p \cos\theta) \geq 2~(m_e c^2 + \frac{m_e}{m_p} m_e c^2)\,,
\eeq where $\gamma_p=E_p/m_pc^2$ is the proton Lorentz factor and $\epsilon_{\gamma}$ the ambient photon 
energy. Thus, for head-on collisions the process is energetically allowed when $\gamma_p \epsilon_{\gamma} > 
m_e c^2$. The maximum energy of the resultant electron (positron) is determined by the kinematics and (for $m_e 
c^2 \ll \gamma_p \epsilon_{\gamma} \ll m_p c^2$) given by (cf. inverse Compton scattering) $E_{e,{\rm max}} = 4 
\gamma_p^2 \epsilon_f$. The cross-section for this process (often referred to as Bethe-Heitler cross-section) is about 
two orders of magnitude higher than the one for photo-meson production, but on average only a small fraction of the 
proton energy ($\sim m_e/m_p$) is lost. This is different to photo-meson production, where a proton transfers on 
average $\sim 13\%$ and more of its energy to the secondary products. As a consequence, photo-meson production 
usually becomes the dominant energy loss channel above the threshold eq.~(\ref{photo-meson-threshold}).\\
A variant of the Bethe-Heitler pair production process in the ergosphere of a maximally rotating, supermassive black 
hole (referred to as Penrose Pair Production) has been suggested already some 30 years ago as an efficient way for 
energy extraction from rotating Kerr black holes:\cite{leiter78,kafatos79} Upon entering the ergosphere, photons 
with energies of some tens of MeV (as might be produced in a hot ADAF inner disk via decay of neutral pions from 
pp-collisions, ref.\cite{mahadevan97b}) could get boosted (blue-shifted) by a factor of up to $\sim 30$ (for extreme 
Kerr black holes) and so approach GeV energies. These photons could then interact with the (low-energy) protons 
moving on marginally stable orbits deep within the ergosphere to produce electron-positron pairs. From the 
threshold condition, minimum photon energies of only $2m_e c^2$ would be required. However, in order to 
become a Penrose Process, the recoiling protons must be injected on negative energy orbits (and pass through 
the event horizon), with the escaping pairs ejected with an energy boost picked up from the rotational energy 
of the Kerr black hole. Kinematically this requires the blue-shifted photon to have an energy comparable to the 
rest mass energy of the proton, giving GeV energies to the escaping electron-positron pairs. While this process 
may in principle be possible, detailed Monte Carlo simulations suggest it to be less important as most of the
pairs apparently do not manage to escape.\cite{williams95}

\subsection{Inelastic proton-proton collisions} 
Relativistic protons and nuclei can produce high energy gamma-rays in inelastic collisions with ambient 
gas.\cite{kelner06} The neutral $\pi^0$-meson (with main $\simeq 99\%$ decay mode into two photons, 
e.g., $p+p \rightarrow p+p + \pi^0$, $\pi^0 \rightarrow  2 \gamma$) provides the main channel of conversion of 
kinetic energy to high energy gamma-rays. For $\pi^0$-production to become kinematically possible, the kinetic 
energy of the bombarding proton must exceed $E_{\rm th} = 2m_{\pi} c^2 (1+m_{\pi}/[4m_p]) \simeq 280$ MeV.
The cross-section and the inelasticity for pp-collisions both depend weekly on the energy, i.e., $\sigma_{pp} 
\simeq 3.4 \times 10^{-26}$ cm$^2$ and total inelasticity $f_{\pi}\simeq 0.5$. The characteristic cooling timescale 
$t_{pp} \sim 1/(\sigma_{pp} n_p f_{\pi} c)$ for a relativistic proton thus becomes
\beq\label{pp-cooling}
 t_{pp} \sim 2 \times 10^7 \left(\frac{10^8 \mathrm{cm}^{-3}}{n_p}\right)~\mathrm{sec}\,,
\eeq where $n_p$ is the ambient gas density. While $t_{pp}$ is only weakly dependent on the energy of the 
proton, the cooling timescale for $p\gamma$-reactions, cf. eq.~(\ref{photomeson}), is expected to decrease 
with (increasing) proton energy (i.e., lower photon energy). Under such conditions, pp-interactions can 
dominate the cooling only below a certain proton energy (cf. also ref.\cite{sikora87} for an application to AGN).\\ 
In the simple delta-functional approximation,\cite{aharonian_a00} the neutral pion takes away a mean fraction 
$f_{\pi^0}\simeq 0.17$ of the kinetic energy of the proton $E_{\pi^0}=f_{\pi^0} E_{\rm p,kin}$, and this energy 
is equally distributed among its decay products. The shape of the photon spectrum is then similar to the 
shape of the parent proton spectrum, but shifted in energy by a factor $f_{\pi^0}$.

\subsection{$\gamma\gamma$-Absorption}
Photon-photon interactions can produce electron-positron pairs ($\gamma + \gamma \rightarrow e^+ + e^-$) and 
thereby lead to a suppression of the VHE photon flux from a source, once the kinematic threshold condition for pair 
production $(h \nu) (h\nu_t) > 2 (m_e c^2)^2/(1-\cos \alpha)$ is satisfied ($\alpha$ being the angle between the 
incident directions of the photons). TeV photons, for example, can thus react (head-on) with infrared target photons 
of energy $\epsilon_t = h \nu_t \sim 0.3$ eV. The electron-positron pairs produced in energetic collisions will be 
highly relativistic and tend to move in the direction of the initial VHE $\gamma$-ray.\\ 
For a fixed photon energy $\epsilon$, the pair-production cross-section $\sigma_{\gamma\gamma}$ is a function 
of target energy $\epsilon_t$:\cite{jauch76} If we denote by $s=E^2/(4 m_e^2 c^4)$ the normalized 
centre-of-mass energy of the photons squared ($E^2=2 \epsilon\epsilon_t [1-\cos\alpha]$), then starting from zero 
at the threshold energy $s=1$, the cross-section rises steeply to a maximum $\simeq (\sigma_T/4)$ at $s\simeq 2$ 
and decreases approximately $\propto 1/s \propto 1/\epsilon_t$ towards higher energies, i.e.,
\beq
 \sigma_{\gamma\gamma}(s) \simeq  \frac{3}{8} \frac{\sigma_T}{s}  
                                                     \left[\ln(4 s)-1\right]
\eeq for $s \gg 1$. In an isotropic background radiation field, TeV photons thus interact most efficiently with infrared 
target photons of energy
\beq\label{gamma_gamma}
     \epsilon_t \simeq 1~\left(\frac{1 \mathrm{TeV}}{\epsilon}\right)~\mathrm{eV}\,.
\eeq The optical depth, characterizing the absorption of a $\gamma$-ray with energy $\epsilon$ moving through 
a target  photon gas of spectral and spatial distribution $n(r,\epsilon_t)$ in a source of size $R$, can then be 
calculated from
\beq
    \tau(\epsilon) = \int_0^R dr \int_{2 (m_e c^2)^2/[\epsilon (1-\cos\alpha)]}^\infty d\epsilon_t~n_{\gamma}(r,\epsilon_t)~
                                  \sigma_{\gamma\gamma}(\epsilon,\epsilon_t,\alpha)~(1-\cos\alpha)\,.
\eeq For a homogeneous source and a quasi-isotropic target photon field, the optical depth becomes\footnote{A full 
analysis would have to employ the appropriate angle-averaged cross-section $\bar{\sigma}_{\gamma\gamma}$, see
e.g. \cite{aharonian_book04}.}
\beq
 \tau (\epsilon) \simeq R \int_{2(m_e c^2)^2/\epsilon}^{\infty} d\epsilon_t ~ n_{\gamma}(\epsilon_t) ~ 
                        \bar{\sigma}_{\gamma\gamma}\,,
\eeq where $n_{\gamma}(\epsilon_t)$ is the differential target photon number density. 
Because the optical depth for a gamma-ray of energy $\epsilon$ is essentially determined by a relatively narrow band 
of target photons (centered at $s=2$), a useful order-of magnitude estimate is given by $\tau(\epsilon)\simeq 2.5 
(\sigma_T/4) R \epsilon_t n_{\gamma}(2\epsilon_t)$.\cite{herterich74}  In terms of the luminosity $L_t$ above energy 
$\epsilon_t \propto 1/\epsilon$, the optical depth may thus be approximated by
\beq\label{optical_depth}
 \tau(\epsilon) \simeq \frac{5 \sigma_T L_t}{128\pi R c \epsilon_t} 
                          \simeq 0.3 \left(\frac{L_t}{10^{40}~\mathrm{erg/s}}\right)\left(\frac{10^{16}~\mathrm{cm}}{R}\right)
                                       \left(\frac{\epsilon}{1~\mathrm{TeV}}\right)\,.
\eeq When applied to AGNs, this suggests that only in highly under-luminous ($L_t \ll L_{\rm Edd}$) and massive ($R
\propto r_s \propto M$) sources, VHE gamma-rays may be able to escape unabsorbed from the vicinity of the central 
supermassive black hole. As an application, let us suppose that the inner accretion disk is of the high-temperature 
($T_e \sim 5 \times 10^9$ K) ADAF type (see \S~\ref{riaf}). The dominant part of the radiation in the hard X-ray regime 
($h\nu \sim k T_e \sim 500$ keV) would then be due to thermal bremsstrahlung. These hard X-ray photons would 
preferentially interact with target photons of energy $\epsilon_t \sim 2$ MeV (cf. eq.~[\ref{gamma_gamma}]). Although 
the disk spectrum would be suppressed at these energies (as the bremsstrahlung spectrum falls of exponentially with 
$\exp[-h\nu/kT_e]$), it may still be strong enough to lead to some non-negligible electron-positron pair production. 
We can roughly estimate the density of pairs so provided by balancing the rate of their creation with the rate of their 
escape:\footnote{I  owe this argument to Marek Sikora.} The pair creation rate [particles/sec] is of the order of 
$\dot{N}_i \sim (4/3) \pi R^3 n_{\gamma}/t_{\rm life}$ where $t_{\rm life} \sim 1/(n_{\gamma} \sigma_{\gamma\gamma} c)$ 
is the characteristic lifetime of an incident photon. The created pairs will escape from the source region at a rate 
$\dot{N}_e \sim 4\pi R^2 n_e c$ (ignoring Compton scattering). Hence we find $n_e \sim n_{\gamma}^2 
\sigma_{\gamma\gamma} R/3$. Using $n_{\gamma} \sim L_t/(4\pi R^2 c \epsilon_t)$, we have 
\beq\label{paircreation}
 n_e \sim \frac{L_t^2 \sigma_{\gamma\gamma}}{48 \pi^2 R^3 c^2 \epsilon_t^2} 
        \simeq 3 \left(\frac{L_t}{10^{40} \mathrm{erg}}\right)^2 \left(\frac{10^{14} \mathrm{cm}}{R}\right)^3
        \left(\frac{2~\mathrm{MeV}}{\epsilon_t}\right)^2 ~[\mathrm{particles\,cm}^{-3}]\,.
\eeq Given the conventional AGN parameter space, one may expect the so-estimated pair density to be comparable 
to the Goldreich-Julian density in a fair number of sources, implying that a substantial part of the parallel electric field 
in the magnetosphere may be screened. This may not be the case, however, for highly under-luminous systems such
as M87 and Sgr~A*.\cite{lev11,mosci11} In terms of the accretion rate, the density ratio of so-created pairs (in a 
two-temperature ADAF) to the GJ value approximately follows\cite{lev11}
\beq
    \frac{n_e}{n_{\rm GJ}} \sim 10^{12}\dot{m}^{7/2} \left(\frac{M}{10^8 M_{\odot}}\right)^{1/2}\,.
\eeq Hence, if the accretion rate would become very small, an additional plasma source (such as cascade formation 
in starved magnetospheric regions) would be needed to establish a force-free outflow (where $n_e \geq n_{\rm GJ}$).\\ 
Note that if a source is sufficiently compact, its VHE $\gamma$-ray emission could well be shaped by the development 
of an electromagnetic cascade (photo-production of pairs, subsequent inverse Compton scattering etc.).\cite{agaronyan84} 
Also, since both TeV $\gamma\gamma$-absorption and photo-meson production, \S~\ref{photo-meson} can interact 
with similar target photon fields, TeV gamma-ray observations could constrain the photo-pion production opacity, in 
particular for sources where rapid VHE gamma-ray variability is observed. When applied to TeV blazars, this suggest 
that the expected flux of high energy neutrinos from $p\gamma$-interactions is much less than the observed VHE 
gamma-ray flux, making them less promising targets for neutrino observations.\cite{levinson06b}

\section{APPLICATIONS}
\subsection{Low-luminous galactic nuclei}
According to the considerations above, nearby under-luminous and non-aligned 'active' galactic nuclei 
emerge as prime candidate sources where non-thermal VHE processes, occurring close to the central black 
hole, may become observable and thereby allow to probe the most violent region of the central engine: If the 
source is sufficiently {\it under-luminous}, VHE gamma-rays may be able to escape significant absorption. 
The source must then, however, be relatively {\it nearby} in order to be within the reach of current or upcoming 
telescope sensitivities. Finally, if the jet is sufficiently {\it misaligned}, the nearby black hole VHE emission may 
not be swamped by Doppler-boosted jet emission. This makes sources such as Sgr~A* or the nearby 
radio-galaxies M87 and Cen~A to interesting candidates.    
 
\subsubsection{The galactic centre source Sgr~A*} 
\begin{figure}[th]
\center
\psfig{figure=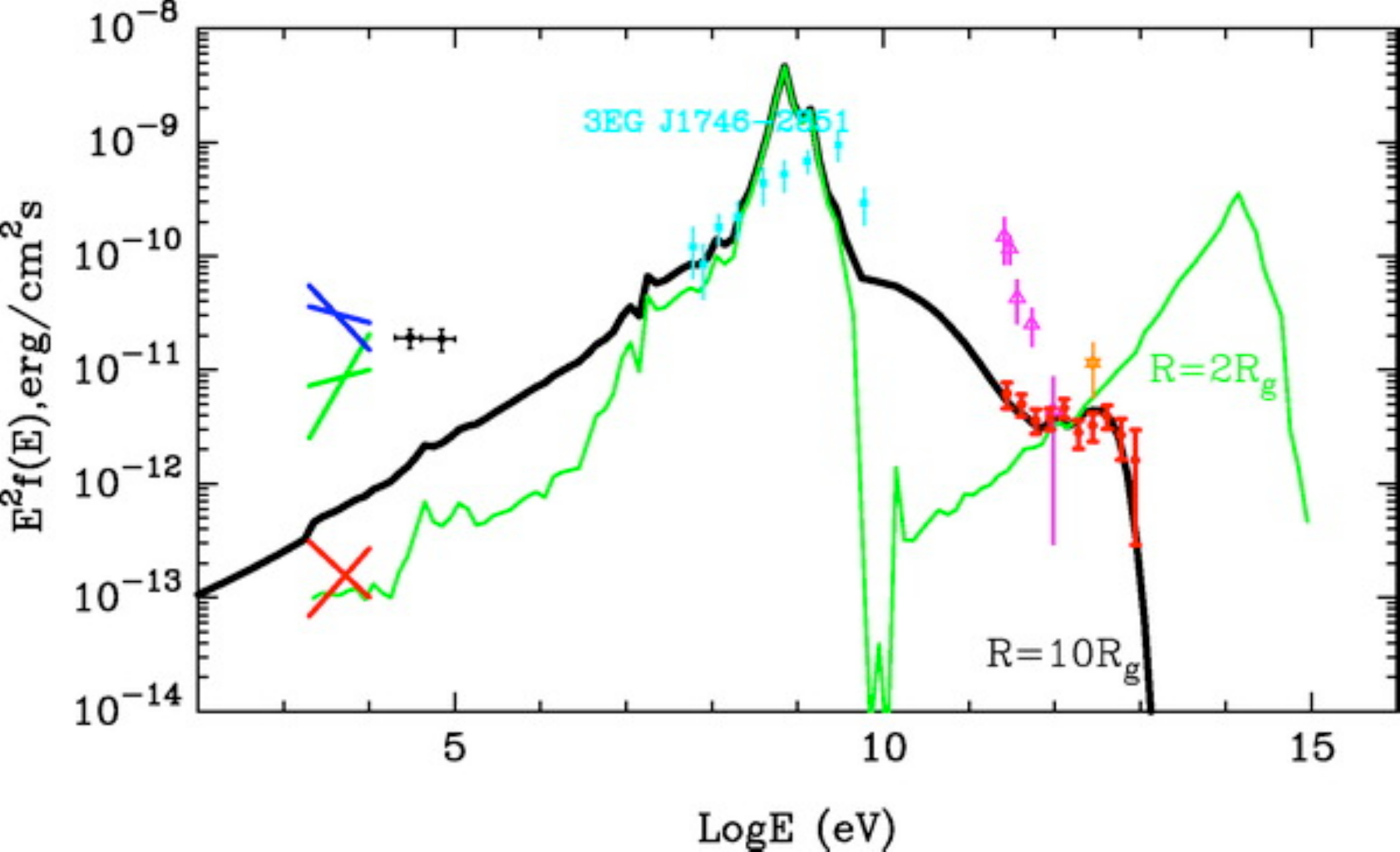, width=10truecm}
\caption{Possible broadband SED for Sgr~A* produced by electron curvature (1st peak) and inverse Compton (2nd
peak) radiation. The thin solid line represents the gamma-ray spectrum due to the accelerated (primary) electrons, 
the thick solid line shows the redistribution of the energy after the passage through the infrared source. Figure 
adapted from Aharonian \& Neronov\protect\cite{aharonian05a} (reproduced by permission of the AAS).}
\label{sgrA-sed}
\end{figure}

Given its proximity, the compact radio source Sgr~A* in the central region of our Galaxy may represent the most 
promising target for studying particle acceleration processes near the event horizon of a supermassive black 
hole. Believed to host a black hole of mass $M_{\rm BH} \simeq (3-4) \times 10^6 M_{\odot}$,\cite{reid09} its 
extraordinary low bolometric luminosity ($\leq 10^{-8} L_{\rm Edd}$) is expected to make the innermost region 
transparent to $\gamma$-rays with energies up to $\sim10$ TeV.\cite{aharonian05a} 
Estimates for the accretion rate in Sgr~A* are in the range $\dot{M} \simeq (0.1-4) \times 10^{-6} 
M_{\odot}/$yr,\cite{baganoff03,nayakshin05,cuadra06} so that $\dot{m}=\dot{M}/\dot{M}_{\rm Edd} \lppr 5 \times 
10^{-5}$. This suggests that the magnetic field in the innermost region could possibly be as high as $B \sim 5 
\times 10^3$ G, cf. \S \ref{mfields}.\\ 
Early VHE observations of the Galactic Centre region by Imaging Atmospheric Cherenkov Telescopes have led to 
the detection of a steady, point-like source of VHE gamma-rays at the gravitational center of our Galaxy, coincident 
with the position of at least three counterparts, the supermassive black hole Sgr~A*, the supernova remnant (SNR) 
Sgr~A East (at a projected distance of $3.7$ pc to Sgr~A*), and the pulsar wind nebula G359.95-0.04 (at a projected 
distance of $0.4$ pc to Sgr~A*).\cite{aharonian04,kosack04,crocker05,aharonian09b,vanEldik09} 
Recent H.E.S.S. results based on high precision pointing, however, now seem to exclude Sgr~A East as the dominant 
source of the observed VHE gamma-rays.\cite{acero10} On the other hand, both G359.95-0.04 and Sgr~A* remain 
promising VHE counterpart candidates, as models exist that could explain the observed emission extending beyond 
$10$ TeV (e.g., see ref.\cite{hinton07} for G359.95-0.04). 
In the case of Sgr~A*, for example, the origin of the observed TeV emission has been related to (i) non-thermal 
processes in the black hole magnetosphere itself,\cite{aharonian05a} (ii) pp-interactions of escaping high energy 
protons in the surrounding dense gas environment,\cite{aharonian05b,liu06,ballantyne07} or possibly (iii) electron 
acceleration in the termination shock of a wind emerging from the innermost parts of the disk.\cite{atoyan04}\\
Suppose that particles could be efficiently accelerated in voltage gaps close to the black hole horizon. Then assuming 
a rapidly spinning black hole, eqs.~(\ref{potential}) and (\ref{gap-maximum}) suggest that protons may possibly be 
able to reach energies up to $\simeq10^{18}$ eV ($\gamma_p \sim 10^9$).\cite{levinson_boldt} If so, then curvature
emission is expected to peak around $\sim10^{10}$ eV, eq.~(\ref{curv_freq}), i.e., well below the TeV domain. 
However, detectable TeV $\gamma$-ray emission may still be possible due to other radiative channels, and at least 
three different scenarios have been proposed:\cite{aharonian05a}\\ 
(1)~{\it Photo-meson interactions:} If protons are able to reach energies of $\sim10^{18}$ eV, they can start to interact
 with infrared soft photons, cf. eq.~(\ref{photo-meson-threshold}). Although the mid-to-near infrared source in Sgr~A* 
is faint with $L_{\rm IR} \sim 10^{35}$ erg/s, it appears sufficiently compact $r_{\rm IR}  \lppr 10^{13}$ cm (e.g., 
refs.\cite{melia_falcke,genzel03,dodds09}) to ensure a high enough seed photon density $n_{\gamma} \sim (10^{12}-
10^{13})$ cm$^{-3}$ for effective interactions with protons. The mean free path of protons through such a photon field, 
however, would be comparatively large, i.e., $\lambda \sim 1/(\sigma_{p\gamma}n_{\gamma} K_p) \sim (10^{15}-
10^{16})$ cm, cf. \S \ref{photo-meson}. This suggests that only a modest fraction $r_{\rm IR}/\lambda \leq 0.01$ 
of the energy in protons can be converted into secondary particles. Thus, in order to account for a VHE flux level of 
$L_{\gamma} \sim 10^{35}$ erg/s,\cite{aharonian04} an injection power up to $\sim 10^{38}$ erg/s in high energy 
protons would be required.\\ 
(2)~{\it Proton-proton interactions:} If protons are not accelerated to energies $\sim 10^{18}$ eV in Sgr~A* (as e.g. 
expected in the case of centrifugal acceleration), the efficiency of photo-meson interactions would be significantly 
reduced. Interactions of energetic protons with the ambient plasma could then become the main channel of VHE 
$\gamma$-ray production. The associated cooling time scale for pp-interactions is of order $t_{pp} \sim 2 \times 
10^7~(10^8\mathrm{cm}^{-3}/n_p)$ sec, cf. eq.~(\ref{pp-cooling}), for typical densities $n_p \sim (10^7-10^8)$ 
cm$^{-3}$ in the inner regions of the accretion flow near Sgr~A*, cf. eq~(\ref{adaf-density}). If the efficiency of 
gamma-ray production is determined by the ratio $e_r$ of accretion time scale  $t_a \sim r/v_r \sim 10^3$ sec 
(with $v_r \sim \alpha v_f$, $v_f$ the free fall velocity) to cooling time scale,\cite{aharonian05a} then the efficiency 
of converting the energy of accelerated protons to secondaries may be as small as $e_r \sim 10^{-5}$. Hence, a 
high injection power of $L_{\gamma}/e_r \sim (10^{39}-10^{40})$ erg/s would be required, very close to what may 
actually be provided by accretion.\\ 
(3)~{\it Inverse Compton scenario:} Upscattering of soft photons by accelerated electrons may provide the most 
economic way to produce VHE $\gamma$-rays. If the infrared source represents the main target field, inverse 
Compton upscattering will proceed in the Klein-Nishina regime, requiring electrons energies $\gppr 10$ TeV to 
account for the observed TeV emission. In order to avoid suppression by dominant electron synchrotron losses, 
particle acceleration must then essentially proceed along the ordered magnetic field. Both, centrifugal and/or 
gap-type particle acceleration are potential candidates to this end. If gap-type particle acceleration would be 
limited by curvature losses, for example, up-scattering could result in VHE $\gamma$-rays with energies 
$E_{\gamma} \sim \gamma_e m_e c^2 \sim (10^{14} -10^{15})$ eV, cf. eq.~(\ref{gap-maximum}). These energetic 
photons will, however, not be able to escape the source, but instead interact with far-infrared photons to produce 
electron-positron pairs. Synchrotron radiation and inverse Compton scattering by these secondary electrons will 
lead to a re-distribution of the initial gamma-ray spectrum. Figure~\ref{sgrA-sed} shows a model calculation of the 
expected spectral energy distribution within such a scenario. 

\subsubsection{The radio-galaxy M87}
\begin{figure}[th]
\center
\psfig{figure=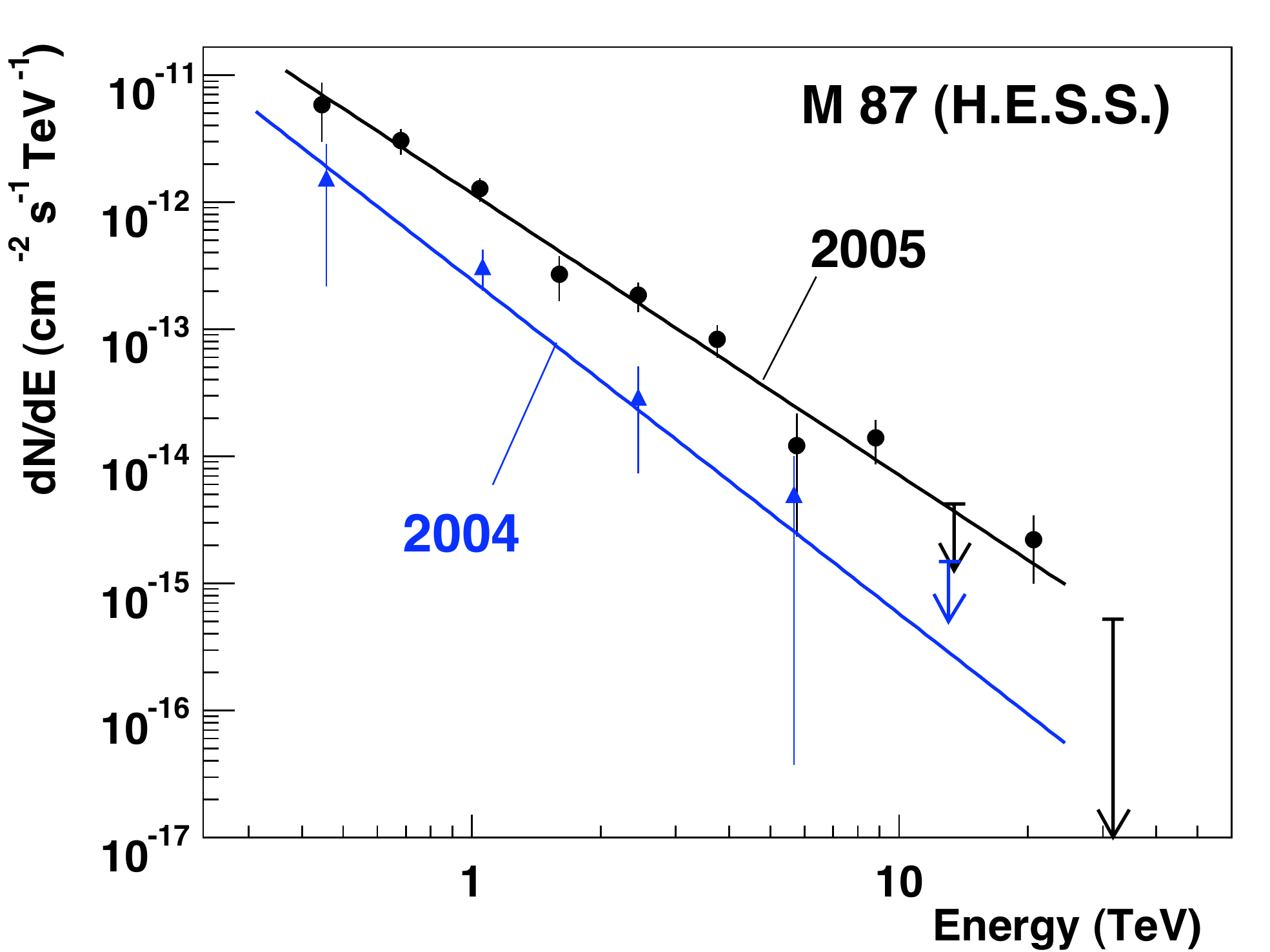, width=9truecm}
\caption{The differential very high energy (VHE) spectra of M87 as obtained by H.E.S.S. during a low (in 2004) 
and a bright (in 2005) state, ranging from $\sim 400$ GeV to $\sim 10$ TeV. The indicated fits give power law 
photon indices of $(\alpha+1)= 2.62 \pm 035$ (2004 data) and $(\alpha+1) = 2.22 \pm  0.15$ (2005 data). No 
variation in spectral shape is found within errors. For comparison: One Crab unit at 1 TeV corresponds to a 
flux of $\simeq 3\times10^{-11}$ s$^{-1}$ cm$^{-2}$. Figure from Aharonian et al.\protect\cite{aharonian06}}
\label{m87_spectrum}
\end{figure}

Located at a distance of $d \sim 16$ Mpc (redshift of $z=0.0043$), the giant elliptical Virgo-cluster galaxy M87 
represent another promising candidate for magnetospheric VHE $\gamma$-ray emission. M87 hosts one of the 
most massive black holes $M_{\rm BH} \simeq (3-6) \times 10^9\,M_{\odot}$ in the Universe\cite{marconi97,gebhardt09} 
and shows a prominent, non-aligned jet detectable from radio to X-ray wavelengths. Despite its huge black hole 
mass, M87 is not a powerful source of radiation: Based on its  estimated total nuclear (disk and jet) bolometric 
luminosity of $L_{\rm bol} \sim 10^{42}$ erg/s or less,\cite{owen00} M87 is highly underluminous with $l_e \leq 3 
\times 10^{-6}$, where $l_e = L_{\rm bol}/L_{\rm Edd}$ and $L_{\rm Edd}$ is the Eddington luminosity. 
This has led to the proposal that M87 is a prototype galaxy, where accretion occurs in an advective-dominated 
(ADAF) mode characterized by an intrinsically low radiative efficiency.\cite{reynolds96,camenzind99,dimatteo03}  
HST observations of M87 have revealed superluminal motion of jet components at $\sim 0.5$ kpc from the central 
black hole, indicative of bulk flow Lorentz factors $\Gamma_b \sim 6$ and a jet orientation of $\theta \sim 19^{\circ}$ 
to the line of sight (ref.\cite{biretta99}; see also ref.\cite{ly07} for larger $\theta$ based on 43 GHz radio observations), 
suggesting that M87 is a non-blazar jet source, characterized by only moderate Doppler factors 
$D=1/[\Gamma_b (1-\beta\cos\theta)] \lppr 3$.\\ 
Despite this, recent TeV observations\cite{aharonian06,albert08,acciari08} have demonstrated that the 
$\gamma$-ray spectrum of M87 extends beyond 10 TeV and is consistent with a relatively hard power-law 
(with spectral index $\alpha \sim 1.2$, where $S_{\nu} \propto \nu^{-\alpha}$), see Fig.~\ref{m87_spectrum}. 
The TeV output is relatively moderate, with an isotropic TeV luminosity of some $10^{40}$ erg/s. Significant 
variability (flux doubling) on time scales of $\Delta t_{\rm obs} \simeq (1-2)$ days has been found, the fastest 
variability observed in any waveband from M87 so far. 
The observed spectral and temporary VHE $\gamma$-ray characteristics of M87 have been proven difficult to 
account for within classical jet models (cf. ref.\cite{rieger09b} for review) and renewed the interest into magnetospheric 
emission models.\cite{neronov07,rieger08a,rieger08b,lev11} This interest has been strengthened by the recently 
detected correlation between radio VLBA (probing scales down to some tens of $r_s$) and TeV $\gamma$-ray 
data during a flaring state of M87 in 2008, see also Fig.~\ref{m87_lc}.\cite{acciari09} There are also indications 
that a simple extrapolation of the Fermi (200 MeV--30 GeV) high energy spectrum under-produces the TeV flaring 
state,\cite{abdo09a} supporting the possible appearance of a new (variable) component at the highest energies.\\
If the observed TeV emission would indeed be due to magnetospheric processes, one may expect the light 
travel time across the source of $\gppr 0.2$ days to provide a lower limit on the possible variability time scale 
at VHE energies. This is obviously close to what can be probed with existing Cherenkov arrays, and is certainly 
within the reach of future instruments like CTA. The observed day-scale variability already implies, e.g., that the 
(generalized) light surface, eq.~(\ref{light-surface}), must be very close to the light cylinder scale of a few $r_g$, 
which in the direct acceleration model \S~\ref{beskin} is equivalent to the requirement of a small longitudinal 
current. The observed, fast variability also excludes, e.g., pp-interactions as dominant channel for VHE 
$\gamma$-ray production in M87.\\
If the accretion rate in M87 is indeed of the order of the Bondi accretion rate $\dot{M} \sim 10^{-3} \dot{M}_{\rm 
Edd}$,\cite{dimatteo03} then possible magnetic field strengths in the vicinity of the black hole could be 
as high as $B \sim 1000$ G, see \S~\ref{mfields}. This would be consistent (if the black hole is sufficiently 
spinning) with poloidal magnetic field strengths required by a Blandford-Znajek-type process (eq.~[\ref{pmax}]) 
to account for the estimated jet kinetic power of a few $10^{44}$ erg/s.\cite{bicknell96,owen00}\\
If electrons are centrifugally accelerated in an ambient photon field of order $L_{\rm bol}$, achievable particle 
energies may be as high as $\gamma_e \sim (10^7-10^8)$, cf. eq.~(\ref{compton}). Similar (and most likely 
even higher) energies might be achievable if efficient gap-type particle acceleration would indeed take place, 
see eq.~(\ref{BHaccel}).\cite{lev11} If the inner disk is of the ADAF-type, this could facilitate upscattering (Thomson 
regime) of (comptonized) sub-mm ADAF disk photons, i.e., those soft photons above the synchrotron peak 
$\nu_p \sim 10^{11}$ Hz, cf. eq.~(\ref{ADAF_peak}). Upscattering would results in emission at $\sim5~(\gamma_e/
10^7)$ TeV energies and can produce a hard photon spectrum close to what has actually been observed.\cite{rieger08b} 
If, on the other hand, the shape of ambient disk photon field would be more of the standard disk-type, then secondary 
inverse Compton cascade emission, initiated by internal absorption, could become influential at TeV energies as 
well.\cite{neronov07}\\
An important question for magnetospheric scenarios is whether TeV photons can escape from the vicinity of the 
central black hole. In principle, TeV photons of energy $\epsilon$ will interact most efficiently with target photons 
in the infrared regime $\epsilon_T \simeq (1\,\rm{TeV}/\epsilon)$ eV, eq.~(\ref{gamma_gamma}). The relevant 
infrared (IR) photon field needs thus to be sufficiently diluted in order to avoid significant pair-absorption. This 
could well happen if the IR emission is dominated by, e.g., emission from the jet (synchrotron emission) and/or 
a dusty torus on larger scales. For M87, the observed nuclear mid-infrared luminosity has been estimated to be 
$L_{\rm IR} \sim 10^{41}$ erg/s.\cite{whysong04} The corresponding optical depth for TeV photons 
would thus be $\tau \sim 3~(10^{16} \mathrm{cm}/R_{\rm IR})$, eq.~(\ref{optical_depth}), assuming the infrared 
emission to be concentrated in a homogeneous source of size $R_{\rm IR}$. Hence, for $R_{\rm IR} \sim$ some 
tens of $r_s$, TeV photons would be able to escape the source. This latter condition is probably difficult to satisfy 
for a standard-type disk where the emission from the innermost part is expected to peak close to the infrared 
regime, cf. \S~\ref{standard_disk}, but it seems well possible for a radiatively inefficient accretion flow. If one 
assumes, for example, that all of the observed nuclear radio to X-ray flux in M87 arises in an ADAF (which may 
appear over-restrictive given its jet), then $\sim 10$ TeV photons would have to be produced on scales $\gppr 5 
r_s$ (for a Kerr black hole) and $13 r_s$ (for a non-rotating black hole) in order to be able to escape the 
source.\cite{wang08} Recent findings, according to which the infrared emission is consistent with optically-thin 
synchrotron emission,\cite{buson09} suggests that the real situation could be even more relaxed. Hence, there 
are good reasons to believe that, at least for M87, TeV $\gamma$-rays may be able to escape from the vicinity 
of its supermassive black hole.\cite{lev11} This in turn suggests that future high-sensitivity instruments like the 
European CTA project could play a particularly important role in testing magnetospheric VHE $\gamma$-ray 
emission scenarios in M87.

\subsubsection{The radio-galaxy Cen~A}
As the closest active galaxy (distance $d\sim 3.4$ Mpc, ref.\cite{ferrarese07}), the FR~I source Centaurus A 
(Cen~A, NGC 5128) seems another promising candidate worth exploring. Being a spectacular example 
of a gas-rich disk galaxy consumed in a merger with a giant elliptical galaxy, Cen~A belongs to the best 
studied extragalactic objects (see, e.g., ref.\cite{israel98}). Radio observations reveal a peculiar morphology 
with a sub-pc-scale jet and counter-jet, a one-sided kpc-jet, two radio lobes and extended diffusive 
emission. VLBI observations suggest that Cen~A is a non-blazar source, its jet being inclined at a rather 
large viewing angle $i \gppr 50^{\circ}$ and characterized by moderate bulk flow speeds $u_j \sim 
0.5$ c.\cite{tingay98,hardcastle03} Its central black hole mass has been inferred to be in the range 
$M_{\rm BH} = (0.5-3) \times 10^8 M_{\odot}$.\cite{marconi06,neumayer07} 

\noindent With an estimated bolometric luminosity output of the order of $L_b \sim 10^{43}$ erg/s,\cite{whysong04} 
 Cen~A is rather under-luminous (although not highly) and believed to be accreting at sub-Eddington rates 
 $\dot{M} \sim 10^{-3}\dot{M}_{\rm Edd}$. The apparent lack of a big blue bump UV feature seems to 
 indicate that its inner disk is not of the standard-type.\cite{marconi01} In fact, the disk contribution 
seems consistent with a hybrid disk configuration where a standard disk is truncated at $r_t$ and replaced by 
an ADAF in the inner regions close to the central black hole (cf. refs.\cite{evans04,pellegrini05,meisenheimer07}). 
If the inner disk in Cen~A would remain cooling-dominated (standard disk), magnetic field strengths close to 
the black hole of order $B \sim 2 \times 10^3$ G might be expected, cf. eq.~(\ref{mfield_rad}). 
If the disk switches  to a radiatively inefficient mode, the characteristic magnetic field strengths may be higher, 
possibly approaching $B \sim 10^4$ G if its black hole mass is at the lower mass end, see eq.~(\ref{mfield_adaf}).

\noindent  The possible association of some of the Pierre Auger Observatory (PAO) measured UHECR events 
above 57 EeV with Cen~A has recently triggered a number of studies analyzing the efficiency of cosmic-ray 
acceleration in Cen~A.\cite{hardcastle09,kachelriess09,osullivan09,rieger09}  
Based on data up to August 2007, the PAO Collaboration initially reported evidence for an anisotropy at the 
99\% confidence level in the arrival directions of cosmic rays with energies above $\sim 6\times 10^{19}$ 
eV.\cite{abraham07} The anisotropy was measured by the fraction of arrival directions that were less than 
$\sim 3^{\circ}$ from the positions of nearby AGN (within 75 Mpc) from the VCV catalog. While that correlation 
has now decreased using the newly available (twice as large) data set, the updated analysis still suggests 
that a region of the sky around the position of Cen~A has the largest excess of arrival directions relative to 
isotropic expectations.\cite{abreu10} If efficient UHE cosmic-ray acceleration would indeed take place in 
Cen~A, this would seem to require the operation of an additional acceleration mechanism beyond 
gap-type particle acceleration, at least in the case of a proton-dominated composition.\cite{rieger09} 
For even in the ideal case where (i) the ordered poloidal field is assumed to be of the order $B\sim 
10^4$ G, (ii) the black hole to be rapidly spinning $a \sim 1$, (iii) almost the full induced electric potential to 
be available for particle acceleration, and (iv) radiative constraints to be negligible, achievable cosmic ray 
energies are not expected to exceed $E \simeq 2 \times 10^{19} Z~a~(M/10^8 M_{\odot})^{1/2} (\dot{M}/10^{-3}
\dot{M}_{\rm Edd})^{1/2}$ eV by much, cf. eq.~(\ref{potential}). Dropping (iv) by taking curvature losses into 
account would reduce achievable proton energies to $\sim10^{19}$ eV, see eq.~(\ref{gap-maximum}). If, on 
the other hand, a centrifugal-type acceleration mechanism would be operative, achievable energies may be 
even smaller, see eq.~(\ref{max-centrifugal-proton}). Acceleration of protons to energies beyond a few times 
$10^{19}$ eV in regular magnetic fields close to the black hole would thus seem to be disfavored, even if 
Cen~A was in the past in more active stage. The situation is much more relaxed in the case of heavier 
elements like iron nuclei, and this may perhaps be consistent with recent PAO indications for an increase of 
the average mass composition with rising energies up to $E\simeq 10^{19.6}$ eV.\cite{abraham10} 

\begin{figure}[thb]
\center
\psfig{figure=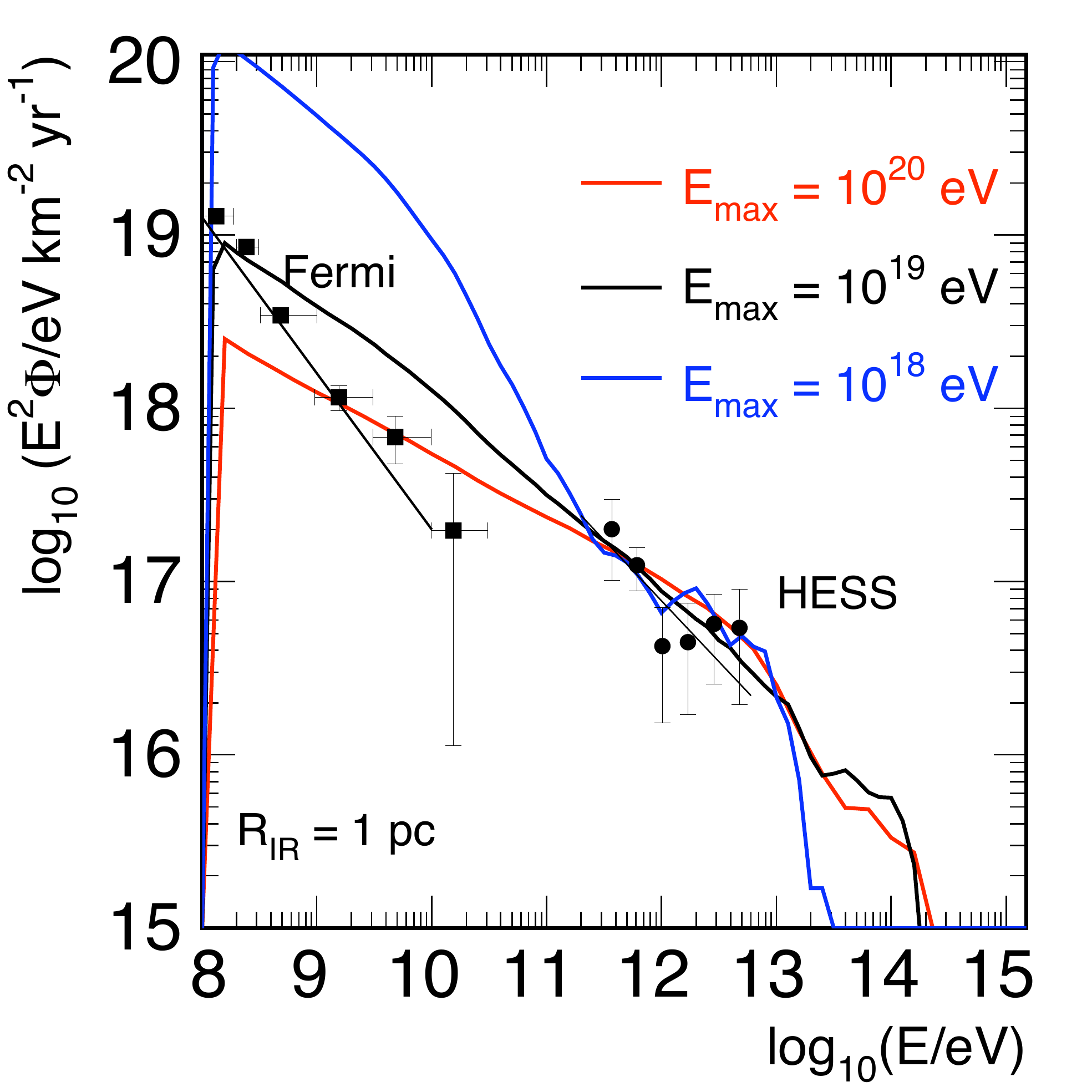, width=6truecm}
\psfig{figure=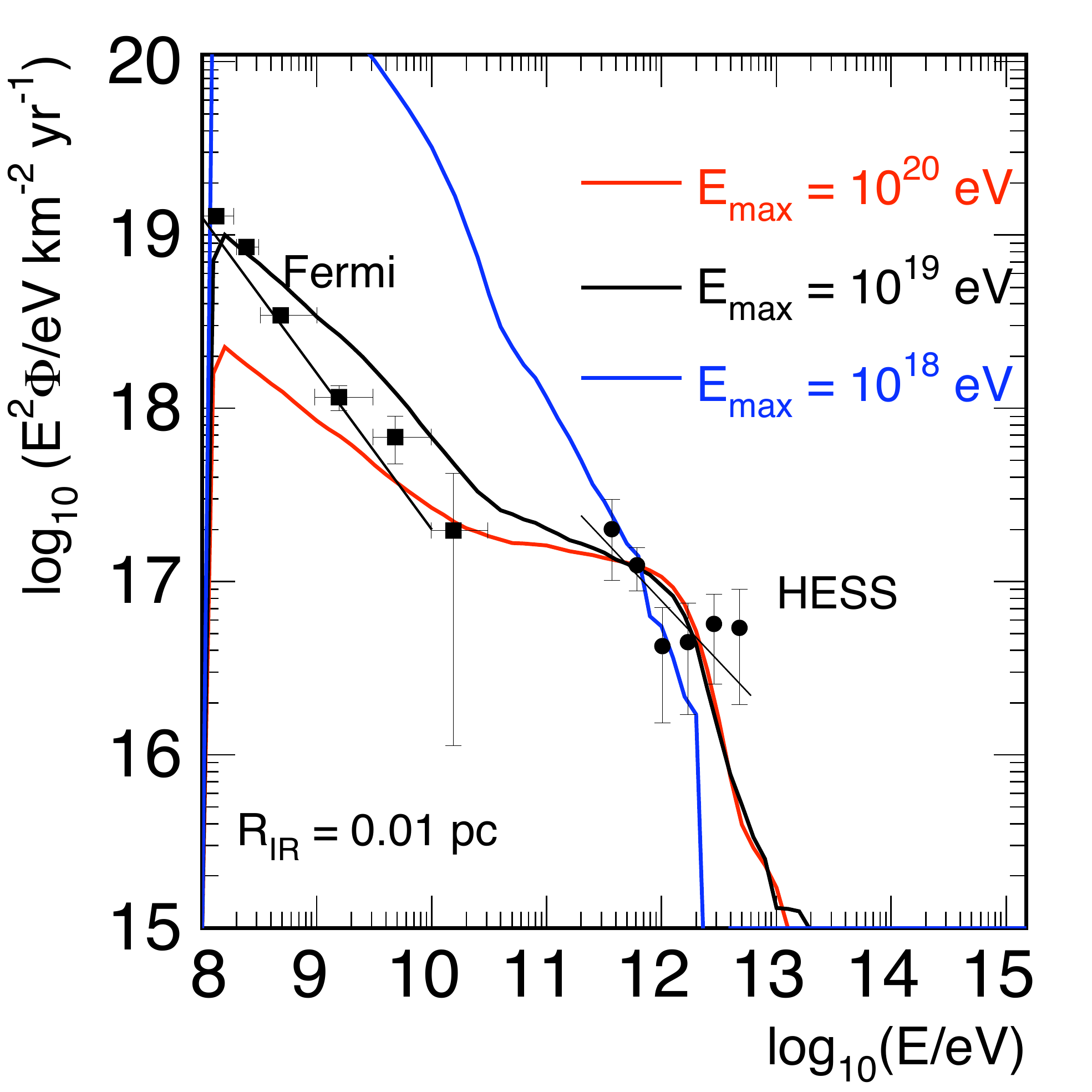, width=6truecm}
\caption{Photon fluxes for Cen~A calculated within a hadronic scenario, assuming p$\gamma$-interactions 
within a strong UV background photon field. Flux suppression by optical depth effects due to the ambient 
IR field becomes apparent towards higher energies. The different lines show the output for different 
maximum proton energies $E_{\rm max}$ assuming either a diffuse or compact IR source, i.e. a spatial IR 
source size of $R_{\rm IR} = 1$ pc and emission cut-off in far IR [left], or a source size 0.01 pc and no 
emission cut-off [right]. In order not to exceed the observed VHE fluxes, proton energies $\gppr 10^{19}$ eV 
seem to be required. Figure adapted from Kachelrie{\ss} et al.\protect\cite{kachelriess10}}
\label{cenA_hadronic}
\end{figure}

\noindent At $\gamma$-ray energies, Cen~A is the only AGN of the non-blazar type detected at MeV 
(COMPTEL) and GeV (EGRET) energies. Its nuclear spectral energy distribution (SED), based on 
non-simultaneous data, appears to be composed of two peaks, one reaching its maximum at several 
times $10^{13}$ Hz and one peaking around $0.1$ MeV.\cite{chiaberge01,meisenheimer07} 
The SED below 1~GeV has been successfully modeled within a simple jet synchrotron self-Compton 
(SSC) framework, assuming Cen~A to be a misaligned BL Lac object and its first SED peak to be due 
to synchrotron emission (ref.\cite{chiaberge01}; for an alternative interpretation, cf. ref.\cite{lenain08}). 
The detection of faint VHE $\gamma$-ray emission (integral flux of $\sim 1\%$ of the Crab Nebula above 
250 GeV) up to $\sim 5$ TeV has been recently reported by the H.E.S.S. collaboration.\cite{aharonian09a} 
The VHE spectrum (from $\sim$ 250 GeV to 5 TeV) can be described by a power law with spectral index 
$\alpha \simeq 1.7 \pm 0.5$. No significant variability has been found in the data set. At lower 
$\gamma$-ray energies (100 MeV to some tens of GeV), Fermi has recently also detected both the 
core and the giant radio lobes of Cen~A.\cite{abdo10a,abdo10b} As it appears, a single population of 
high energy particles seems unable to account for the core flux observed by Fermi and H.E.S.S. In fact, 
a simple extrapolation of the Fermi (power law) spectra would tend to under-predict the flux at TeV 
energies.\cite{abdo10b,raue09} This could possibly indicate an additional contribution to the VHE domain 
beyond the conventional SSC jet emission, emerging at the highest energies. Obvious candidates within 
this context could be: (i) electron inverse-Compton (IC) emission powered by magnetospheric processes 
(centrifugal acceleration of electrons) in an ADAF environment,\cite{rieger09} (ii) IC pair cascade emission 
in the radiation field of a standard disk,\cite{sitarek10} provided the magnetic field is weak enough for 
synchrotron cooling to be suppressed (cf. also ref.\cite{orellana09}), and (iii) hadronic (p$\gamma$) 
interactions in a standard-type disk environment, provided protons can achieve $\gppr 10^{19}$ 
eV.\cite{kachelriess10} These approaches are complementary in that the leptonic scenario (i) requires 
a rather low ambient IR photon field close to the central source, whereas the leptonic model (ii) or the 
hadronic model (iii) are dependent on rather large UV and/or IR background fields. An exemplary 
output of model (iii) is shown Fig.~\ref{cenA_hadronic}.

\noindent Let us suppose that TeV $\gamma$-rays could be indeed produced relatively close to the central 
black hole in Cen~A. Then given its putatively small source scale and relatively high nuclear mid-infrared 
brightness (isotropic $\sim 6 \times 10^{41}$ erg/s at $h\nu \sim 0.15$ eV, with evidence for an exponential 
cut-off towards higher frequencies, see, e.g., refs.\cite{whysong04,meisenheimer07}), efficient escape of these 
VHE photons might appear much more challenging for Cen~A than for M87, cf.~eq.(\ref{optical_depth}). 
However, most of the nuclear IR emission may well arise on larger scales (note that the mid-IR emission is 
unresolved on scales $\sim 0.2$ pc) and thereby allow sufficient dilution. In fact, the observed nuclear mid-IR 
emission is commonly believed not to be produced closed to the black hole, but to be dominated by a 
non-thermal (non-isotropic!) synchrotron jet component at a distance of $\gppr 0.03$ pc (cf. 
refs.\cite{chiaberge01,meisenheimer07,prietro10}) and/or a (quasi-isotropic) dusty torus on scales $\sim 0.1$ 
pc or larger (refs.\cite{whysong04,radomski08}). If this would be indeed the case, then 
photons with energies of a few to several TeV might be able to escape unabsorbed, while absorption 
features should become apparent towards higher energies. This could be tested with future high-sensitivity 
VHE $\gamma$-ray instruments and help to disentangle the site of the VHE $\gamma$-ray production 
and the origin of the observed nuclear mid-IR emission. 
 
\subsection{Passive black holes as TeV $\gamma$-ray and UHE cosmic-ray sources}
If supermassive black holes would indeed be efficient UHE cosmic-ray accelerators, one may expect this to 
be accompanied by a significant TeV $\gamma$-ray output due to curvature radiation, e.g. ref.\cite{levinson09}. 
Observational constraints on the latter may then allow to impose physical constraints on the first. Perhaps 
the most obvious candidates for efficient gap-type particle acceleration are massive, but very weakly-emitting 
(dormant or "non-active") sources:

\subsubsection{The Fornax cluster galaxy NGC 1399}
Consider, for illustration, the giant elliptical Fornax cluster galaxy NGC~1399 at a distance of $\sim 20$ 
Mpc, already mentioned in \S~\ref{BH_mass}. Believed to host a supermassive black hole of mass $\sim
5 \times 10^8 M_{\odot}$,\cite{houghton06,gebhardt07} it is well known for its low nuclear emissivity at 
all wavelengths. NGC 1399 exhibits weak FR~I type radio activity (at a level of $\sim 10^{39}$ erg/s; see
ref.\cite{killeen88}), but little evidence for significant non-thermal jet emission at higher energies. 
Based on an analysis of jet-induced cavities in the surrounding X-ray emitting medium on kpc-scale, a 
jet kinetic power of $P_{\rm BZ}\sim 10^{42}$ erg/s has been inferred.\cite{shurkin08} If this jet power 
would be provided by a Blandford-Znajek-type (BZ) process, then ordered poloidal magnetic field 
strengths close to the black hole of the order of $B_p\sim 300/a$ G (with $a$ the spin parameter) are 
to be expected, cf. eq.~(\ref{pmax}).

\noindent Estimates for the (ADAF-type) accretion rate based on Chandra observations, on the other hand, 
seem to suggest $\dot{M} \sim 0.1~\dot{M}_{\rm Bondi} \sim 2\times 10^{-4} \dot{M}_{\rm Edd}$ 
(see refs.\cite{loewenstein01,narayan02}) and would therefore imply (disk) magnetic field strengths (eq.~[\ref{mfield_adaf}]) 
$B_a \sim 10^3$ G, compatible with the one inferred from the observed jet power. Thus, if in addition the full 
electric potential would be available for particle acceleration (i.e., $a \sim 1$, $h\sim r_g$), electrons/protons 
might be able to reach maximum Lorentz factors $\sim10^{10}$, i.e. protons may reach ultra-high energies of 
$\sim10^{19}$ eV, cf. eqs.~(\ref{potential}) and (\ref{gap-maximum}). This would yield curvature emission in the 
$\sim(0.1-1)$ TeV regime (eq.~[\ref{max_curv_photon}]). An important consequence of this is that a system 
containing a rapidly spinning black hole (such that it  can be an efficient UHECR accelerator) should also 
emit curvature TeV photons, provided (i) vacuum breakdown does not occur and (ii) the TeV photons are able 
to escape.\cite{levinson00} In the case of NGC 1399 the later seems likely to be the case (cf. ref.\cite{pedaletti11}).

\noindent In order to sustain the observed BZ jet power, a current $I\simeq (P_{\rm BZ}/\Delta Z_s)^{1/2} 
\sim 10^{26}~\mathrm{statampere} \simeq 3 \times 10^{16}$ A must flow through the magnetosphere. 
This would require an injection rate of $dN_e/dt \sim I/Q \sim 2 \times 10^{35}$ particles/s and imply a 
minimum charge density $n_e \sim N_e/V \sim (dN_e/dt)/(r_g^2 c) \sim 10^{-3}$ particles/cm$^3$. If 
this cannot be provided by annihilation of MeV disk photons, an additional plasma source would be 
needed to establish a BZ-type jet.

\begin{figure}[th]
\center
\psfig{figure=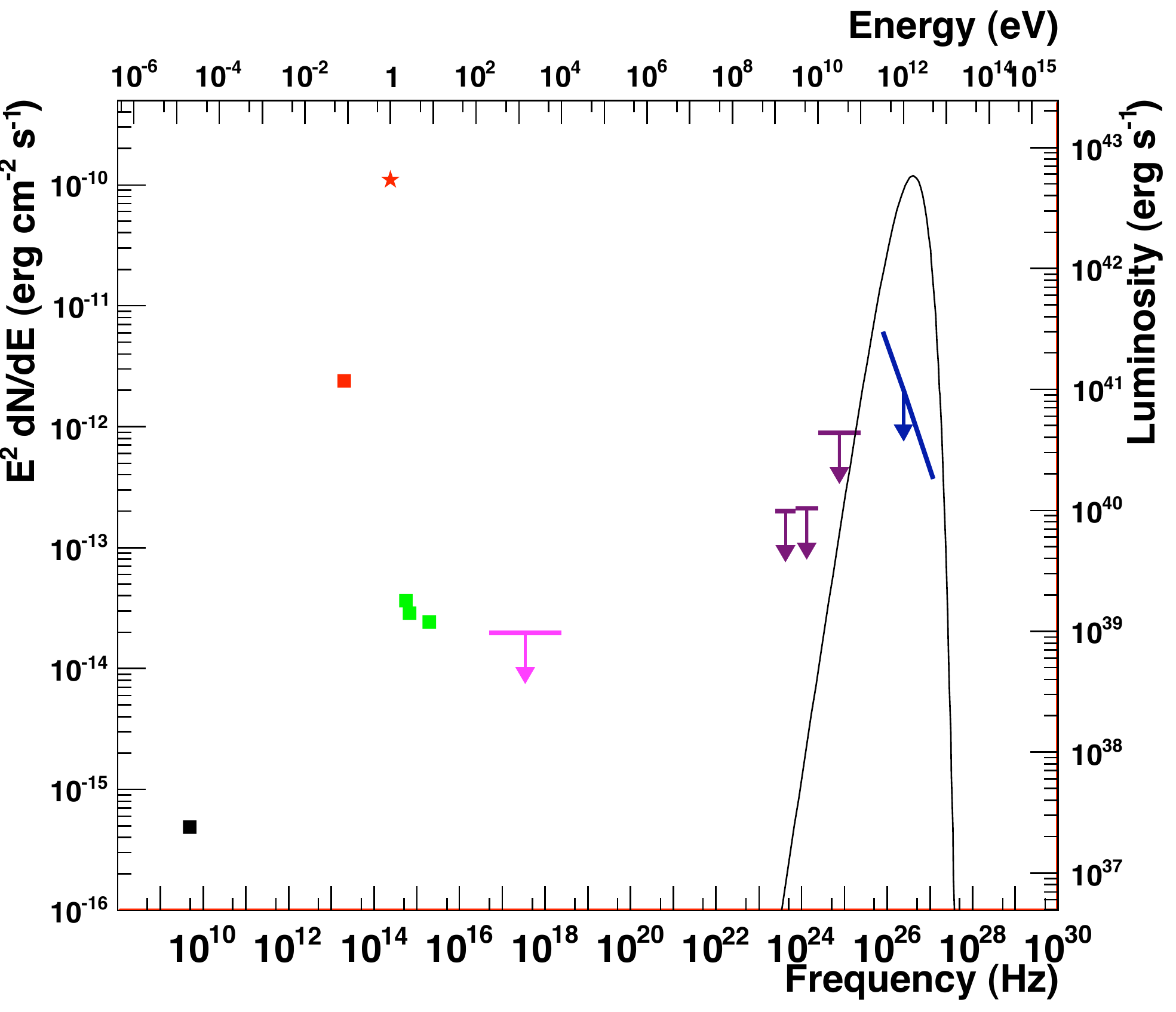, width=10truecm}
\caption{The spectral energy distribution (SED) of NGC~1399, compiled from non-simultaneous public 
data (VLA radio, ISO IR, HST, Chandra, Fermi and HESS). Above the optical regime, only upper limits 
are available. The thin black line represents a toy curvature spectrum assuming a fully developed gap. 
Figure from Pedaletti et al.\protect\cite{pedaletti11}}
\label{ngc1399_sed}
\end{figure}

\noindent The accretion rate in NGC~1399 seems to be close to the critical value $\dot{m}_{\rm crit} 
\sim 3 \times 10^{-4}$ where annihilation of MeV photons in an ADAF could lead to an injection of seed 
charges of density $n_e$ comparable to the Goldreich-Julian value $n_{\rm GJ} \simeq 7 \times 10^{-4}
(B/300 \rm{G})$ cm$^{-3}$, cf. ref.\cite{lev11} If the accretion rate would be sufficiently high, i.e. $\dot{m} > 
\dot{m}_{\rm crit}$ ensuring $n_e >n_{\rm GJ}$, a substantial part of the electric potential may 
be screened (i.e., $h < r_g$ likely) so that efficient gap-type particle acceleration becomes suppressed. 
The anticipated curvature VHE output would be of the order of (cf. eq.~[\ref{curvature_output}]) $L_c 
\sim P_{\rm BZ}~(h/r_g)^2$. For $(h/r_g)^2 \lppr 0.1$, this would be consistent with the VHE upper limit 
$L_{\gamma}(>200~\rm{GeV}) < 9.6 \times 10^{40}$ erg/s, imposed by H.E.S.S. observations of 
NGC~1399, see Fig.~\ref{ngc1399_sed}.\cite{pedaletti11} This would further reduce the available 
potential (eq.~[\ref{gap_potential}]), limiting achievable proton energies to $\ll 10^{19}$ eV. If, on the 
other hand, the accretion rate would be sufficiently small ($\dot{m} < \dot{m}_{\rm crit}$ implying $n_e <
n_{\rm GJ}$), fully developed gaps ($h \sim r_g$) may exist, but an additional plasma source (such as 
cascade formation in starved magnetospheric regions, ref.\cite{lev11}) would be needed to ensure a 
force-free outflow. The gap would then emit VHE $\gamma$-rays with a total luminosity of order 
$L_c \sim (n_e/n_{\rm GJ})~P_{\rm BZ}$. For $(n/n_{\rm GJ})<0.1$ this would again be consistent 
with the VHE upper limits from H.E.S.S. and Fermi observations. 

\subsubsection{Quasar remnants as UHECR accelerators}\label{UHE_sources}
Provided screening is not effective (and the spin parameter high enough), acceleration of cosmic-rays 
to energies $\geq 10^{19}$ eV may become possible. This could perhaps be the case in inactive quasar 
remnants, harboring rapidly spinning, supermassive ($>10^9 M_{\odot}$) black holes.\cite{boldt99} 
The scenario envisaged is then one, where the black hole accelerator is not always operational 
in the normal, BZ-jet producing mode (associated with vacuum breakdown), but where (at least) 
sporadically the full gap size becomes available for the acceleration of a small number of protons. 
Primary target sources would thus be those that show little evidence for jets as well. While this may not 
work for NGC 1399 for reasons given above, Boldt \& Ghosh\cite{boldt99} initially suggested a number 
of nearby massive dark objects that could potentially be efficient UHE ($\geq 10^{20}$ eV) proton 
accelerators. 
However, when curvature losses are fully accounted for,\cite{levinson00} most of these nearby sources 
turn out not to be capable of doing so, see eq.~(\ref{gap-maximum}) (this seems also to apply to the 
revised list in ref.\cite{boldt00}). In fact, there is only a relatively small region of the parameter space (black 
hole mass, magnetic field strength) where one may expect proton acceleration to $\geq 10^{20}$ to be 
marginally possible. For this parameter region, however, the total power of the accompanying VHE 
radiation would appear to exceed the total power emitted in UHE cosmic rays by a factor of a hundred or 
more, making such hypothetical sources no longer quiescent in the TeV regime.\cite{neronov09} 
Given current evidence, it seems thus rather unlikely that (nearby) massive quasar remnants are 
promising candidates for efficient (gap-type) acceleration of protons to energies beyond $5 \times 10^{19}$ eV, 
even if energy losses due to photo-pion production, cf. ref.\cite{boldt00}, would be neglected. As the constraint 
imposed by curvature losses is only mildly dependent on the charge number, heavier nuclei such as, e.g. 
iron Fe, might in contrast be able to reach energies $\gppr 5 \times 10^{20}$ eV in massive remnants. 
The propagation of such nuclei will, however, be significantly influenced by the interactions with the 
intergalactic radiation fields through photo-disintegration, with the particle being stripped from all its 
constituents nucleons after distances of a few Mpc only.\cite{khan05,hooper07}

\section{Conclusions}
Non-thermal processes occurring in the vicinity of supermassive black holes may allow new insights 
into the behavior of matter and energy under extreme conditions. The detection of, e.g., rapid variability 
on timescales of  $\sim r_s/c$ and unusual high-energy emission signatures can provide an important 
diagnostic tool towards this end. Based on this, VHE observations during the last years have demonstrated 
the capacities of the new generation of VHE gamma-ray instruments (e.g., MAGIC II, HESS II, VERITAS 
and FERMI) to substantially improve our knowledge of the activity center in nearby, underluminous AGNs 
and to trigger new theoretical developments that could serve as benchmark for future VHE arrays like 
CTA. Detailed theoretical modelling of non-thermal particle acceleration and radiative numerical simulations 
may soon become indispensable in order to fully take advantage of the improved observational capacities.

\section*{Acknowledgement}
I am very grateful to Felix Aharonian, Vasily Beskin, Christian Fendt, Amir Levinson and Zaza Osmanov 
for helpful comments on the manuscript. Constructive comments by the anonymous referee are also 
gratefully acknowledged.

\end{document}